\documentclass{aa}
\input epsf
\usepackage{graphicx}
\usepackage{natbib}

\begin{document}

\sloppypar

%

\title{Boundary layer, accretion disk and X-ray variability in the
luminous LMXBs.}


   \author{M.Gilfanov\inst{1,2}, M.Revnivtsev\inst{1,2}, S.Molkov\inst{2}}

   \offprints{gilfanov@mpa-garching.mpg.de}

   \institute{Max-Planck-Institute f\"ur Astrophysik,
              Karl-Schwarzschild-Str. 1, D-85740 Garching bei M\"unchen,
              Germany,
        \and   
              Space Research Institute, Russian Academy of Sciences,
              Profsoyuznaya 84/32, 117810 Moscow, Russia
            }
  \date{}

        \authorrunning{Gilfanov et al.}
        \titlerunning{Boundary layer, accretion disk and  variability 
in LMXBs}
 
   \abstract{
Using Fourier frequency resolved X-ray spectroscopy  we study
short term spectral variability in luminous LMXBs.
With RXTE/PCA observations of 4U1608--52 and GX340+0 on the
horizontal/normal branch of the color-intensity diagram we show that
aperiodic and quasiperiodic variability  on  $\sim$ sec -- msec
time scales  is caused primarily by variations of the luminosity of
the boundary layer.
The emission of the accretion disk is  less variable on
these time scales and its power density spectrum follows $P_{\rm
disk}(f)\propto f^{-1}$  law, contributing to observed flux
variation at low frequencies and low energies only. 
The kHz QPOs have the same origin as variability at lower frequencies, 
i.e. independent of the nature of the "clock", the actual luminosity 
modulation takes place on the neutron star surface,

The boundary layer spectrum remains nearly constant in the course of 
the luminosity variations  and is represented to certain accuracy by
the  Fourier frequency resolved spectrum. In the considered
range $\dot{M}\sim (0.1-1) \dot{M}_{\rm Edd}$ it depends weakly on
the global mass accretion rate and in the limit $\dot{M}\sim
\dot{M}_{\rm Edd}$ is close to Wien spectrum with $kT\sim
2.4$ keV (in the distant observer's frame).
The spectrum of the accretion disk emission is significantly softer and
in the 3--20 keV range is reasonably well described by a relativistic
disk model with a mass accretion rate consistent with the value
inferred from the observed X-ray flux. 

   }

   \maketitle

%

\section{Introduction}
\label{sec:intro}

It is commonly accepted that in non-pulsating neutron star X-ray binaries
the magnetic field of the neutron star is  weak enough and the accretion
disk can extend close to the surface of a neutron star. If the neutron star
rotation frequency is smaller than the Keplerian frequency at the inner edge
of the  disk, a boundary layer will be formed near the surface of the
neutron star in which the accreting matter decelerates and spreads over
star's surface \citep{ss86,kluzniak,inogamov99,popham01}. 
For a non-rotating neutron star, in Newtonian approximation  half of
the energy release due to accretion  would take place in the
boundary/spread layer.   
The effects of the general relativity can  increase this fraction,
e.g. up to $\sim 2/3$  in the case of a neutron star with radius
$R_{\rm NS}=3 R_{\rm g}$ \citep{ss86,sibg00}. 
Rotation of the neutron star and deviations of the space-time geometry
from Schwarzschild metric further modify the fraction of the energy
released on the star's surface. 
Consequently, a luminous spectral component, corresponding to the  boundary 
layer emission is expected be present in the X-ray spectrum of a
neutron star X-ray binary

X-ray observations of neutron star LMXBs in the high luminosity state
reveal rather soft composite X-ray spectra.  Based on theoretical
expectations they are usually represented as a sum of two components
attributed to the optically 
thick emission of the accretion disk \citep{ss73} and of the boundary
layer/neutron star  surface \citep{mitsuda84, white88}.
The spectra of these two components are rather similar 
to each other and decomposition of the X-ray emission into the boundary
layer and accretion disk components is often ambiguous,
especially when based on the spectral information alone. 
Not surprisingly, the best fit parameters derived from the data of different
instruments and, correspondingly, the inferred values of the physically
meaningful quantities  are often in 
contradiction to each other \citep[e.g.][]{mitsuda84, disalvo01, done02}.
A robust and sufficiently model independent  method of
separating the boundary layer and disk emission is of interest.

\citet{mitsuda84} and \citet{mitsuda86} studied pattern of spectral
variability on the timescale of $\sim 10^3$ sec in luminous LMXBs and
found that the observed spectra can be represented as a sum of two
components, having drastically different variability properties: a
strongly variable $\sim 2$ keV nearly blackbody component and a stable
softer component. They interpreted the hard and soft components as emission
from the neutron star surface and from the optically thick accretion disk
respectively and concluded that the optically thick disk is stable on the
timescales considered. The latter conclusion is in accord with the finding
of \citet{chur01}, who showed that the same is true in the high luminosity
state of the black hole binary Cyg X-1. On  time scales $\la 10^2$ sec
the fractional amplitude of variations of the disk emission is at least an
order of magnitude lower than that of the hard Comptonized component. 
These results indicate that a low level of variability might be an intrinsic
property of the optically thick accretion disk, independent of the nature 
of the compact object.  

\begin{figure}
\includegraphics[width=0.5\textwidth]{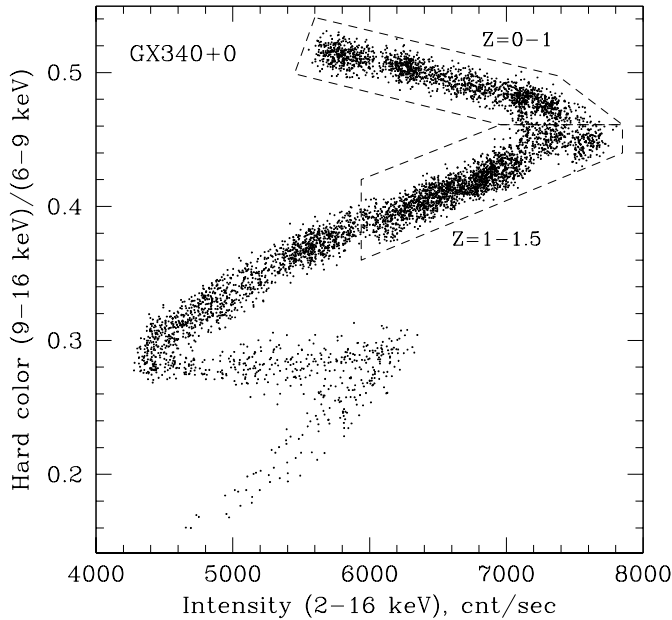}
\caption{The color--intensity diagram of GX340+0. Dashed line polygons
show the regions at the  Horizontal and upper half of the Normal
Branch used for the frequency resolved analysis.  
\label{fig:cid}} 
\end{figure}

\begin{figure}
\includegraphics[width=0.5\textwidth]{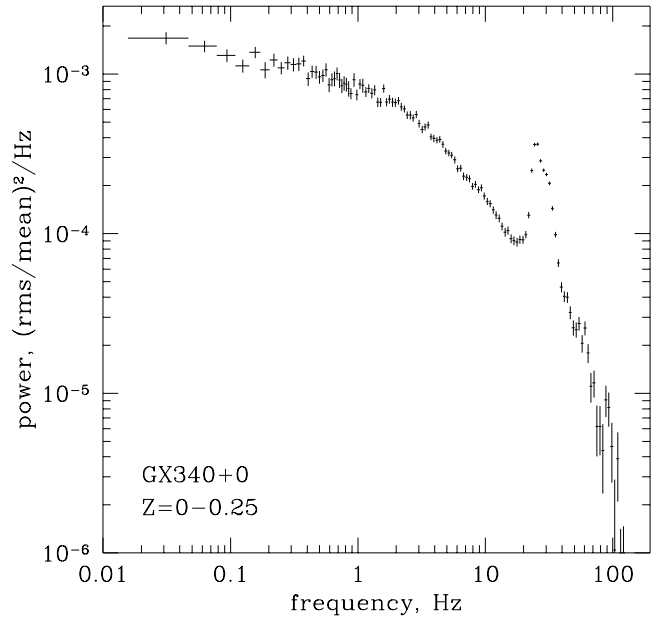}
\caption{Power spectrum of GX340+0 at the upper part of the Horizontal
Branch of the color-intensity diagram.
\label{fig:pds_gx340}}
\end{figure}

As is well known \citep[see][for review]{vdk86,vdk00}, aperiodic
variability of X-ray flux from LMXBs can be 
broadly divided into two main phenomena -- continuum noise and
quasi-periodical oscillations (QPO) with frequencies ranging from several
mHz to more than  a thousand Hz \citep[e.g.][]{has_vdk_89, vdk00,
mikej_mhz}.  
\citet{mitsuda84} studied the difference between the spectra
averaged at different intensity levels -- that restricted the range of
accessible time scales to $\ga 10^3$ sec.
In this paper we will exploit the technique of Fourier frequency resolved
spectroscopy \citep{freq_res99} to study spectral variability of luminous
LMXBs on a broad range of time scales, including kHz QPO.

As defined in \citet{freq_res99}, the Fourier frequency resolved spectrum
is the energy dependent rms amplitude in a selected frequency range,
expressed in absolute  (as opposite to fractional) units. 
A similar approach was used by \citet{mendez0614} to study the energy
spectrum of kHz oscillations in  4U0614+09.
One of it's advantages 
over simple fractional rms--vs.--energy dependence is the possibility to use
conventional (i.e. response folded) spectral approximations in order to
describe the energy dependence of aperiodic variability.  
One should keep in mind, however, that the interpretation of the frequency
resolved spectra often is not straightforward and might strongly depend on a
priori assumptions. Nevertheless, several applications of this technique to
variability of black hole binaries gave meaningful results 
\citep[e.g.][]{freq_res99, gilfanov_cygx1}.

\begin{figure*}
\hbox{
\includegraphics[width=0.5\textwidth]{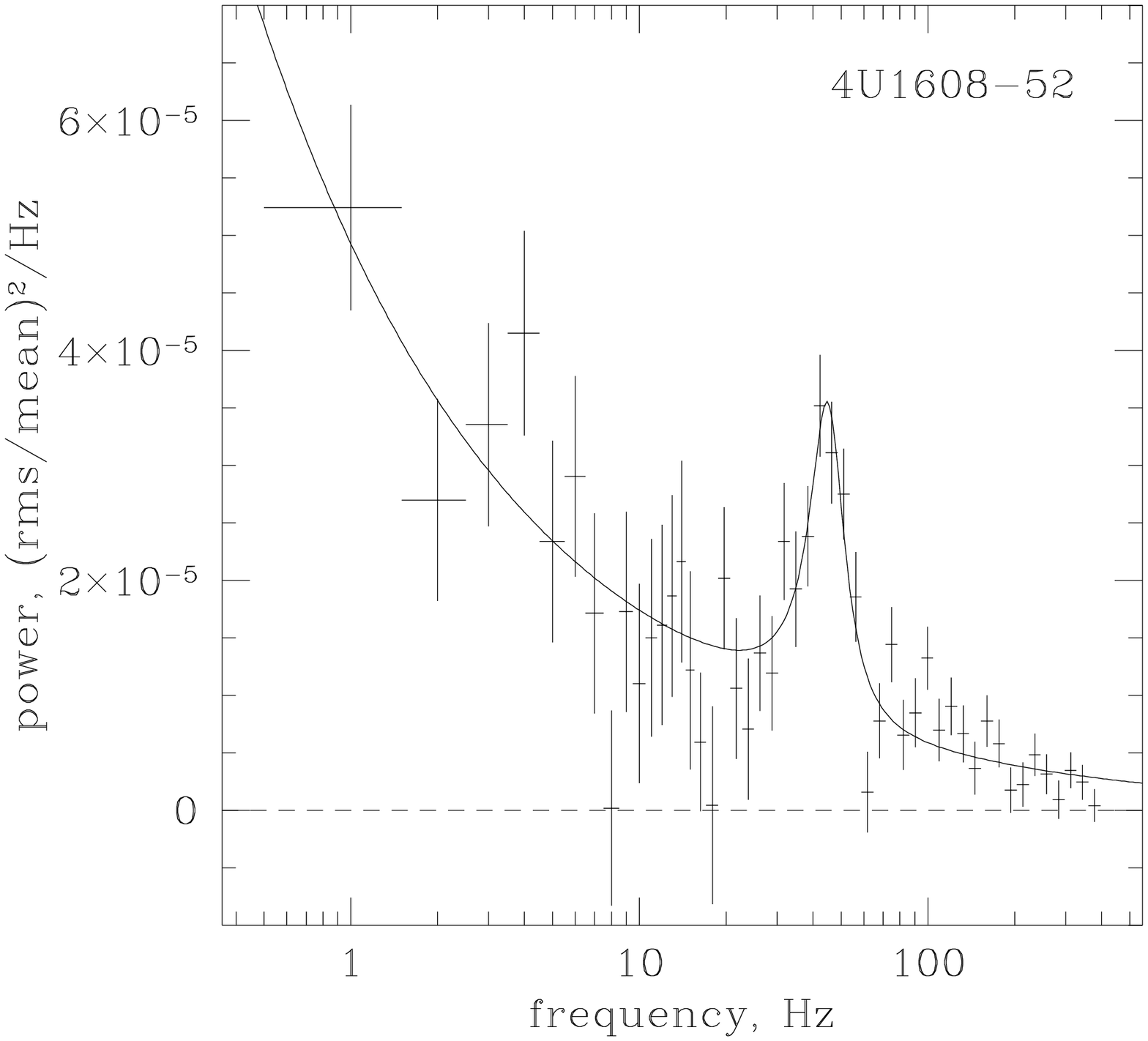}
\includegraphics[width=0.5\textwidth]{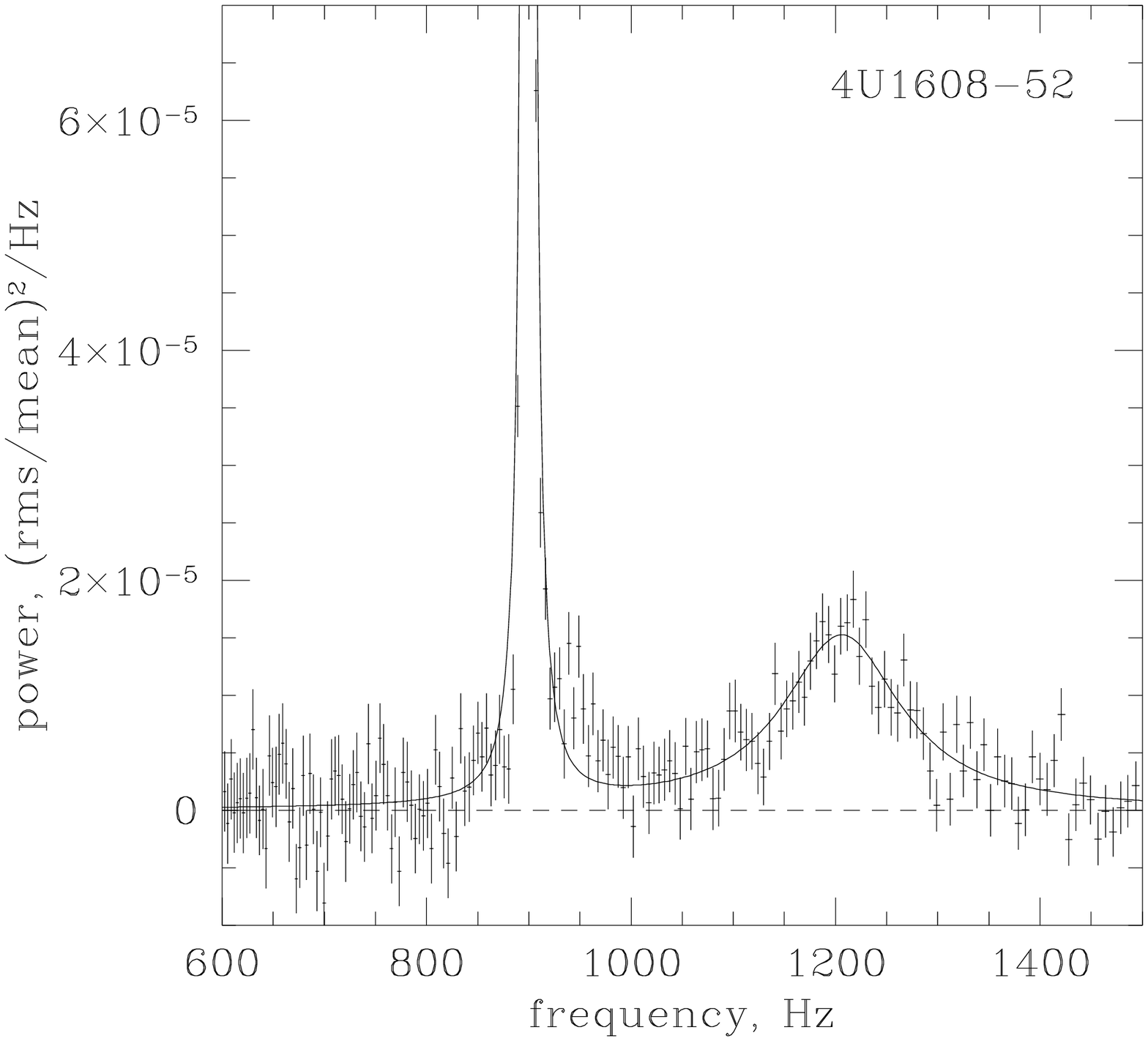} 
}
\caption{The low ({\em left}) and high ({\em right}) frequency parts of the
power spectrum of 4U1608 averaged over all data used for 
analysis. The power spectrum of the high frequency part, showing two kHz QPO
peaks was obtained by ``shift-and-add''  method.
\label{fig:pds_1608}} 
\end{figure*}

The simplest situation, when frequency resolved spectra can be
easily interpreted is illustrated by the following example.
Consider a two-component spectrum, in which one component is stable
and the normalization of the other varies while it's spectral shape is
unchanged. In this case the shape of the frequency resolved spectrum would
not depend on Fourier frequency and would be identical to the spectrum
of the variable component. The spectrum of the non-variable component
could, in principle, be  determined by subtracting the frequency resolved
spectrum from the average spectrum with appropriate renormalization. 
Importantly, in this example the X-ray flux in all energy
channels  will vary coherently and with zero time/phase lag between
different energies. Presence of significant phase lag  and/or Fourier
frequency dependence of the frequency resolved spectra would indicate
that a more complex pattern of spectral  variability is taking place
and interpretation is then less obvious. With few exceptions\footnote{
\label{page:cygx2_fotnote}
$\sim 150\degr$ phase lag was detected in
the normal branch QPO of Cyg X-2 with the pivot energy $\sim 5$
keV, but  no such lags were found in a similar spectral state of 
Sco X-1 \citep{dieters00} 
},
phase lag between light curves in different energy bands in luminous  
LMXBs (e.g. Sco X-1, GX5-1, 4U1608--52, 4U0614+091 etc.)
is usually small, $\Delta\phi\la{\rm few}\times 10^{-2}$, 
coherence is consistent with unity,  \citep[e.g.][]{vaughan94, 
vaughan99, dieters00} and the fractional rms-vs-energy dependence
similar at different Fourier frequencies \citep{vdk86}.
This suggests, that Fourier frequency resolved 
spectral analysis can be applied and its interpretation is
sufficiently straightforward and model-independent.

The structure of the paper is as follows. We briefly describe the
data in Sect.~\ref{sec:data}. In Sect.~\ref{sec:freqres_data} we
present the results of the observations, show that the frequency
resolved spectra do not depend upon the Fourier frequency, and
constrain the time   lags between different energies.
The initial observational results are summarized in
Subsect.~\ref{sec:obs_sum}. In Sect.~\ref{sec:freqres_theory} we
show that a particularly simple form of the spectral variability is
required in order to satisfy the observational constraints -- the flux
variations in different energy channels must be related by a simple
linear transformation.  
In Sect.~\ref{sec:freqres_applications} we compare the expected
spectra of the  disk and boundary layer emission with the frequency
resolved spectra. We show that the observed aperiodic and
quasiperiodic variability is primarily caused by variations of the
luminosity of the boundary layer and its energy spectrum can be
represented, to certain accuracy, by the frequency resolved spectra.
In Sect.~\ref{sec:discussion} we discuss the boundary layer emission
spectrum, it's dependence on the mass accretion rate and implications
of these results for  disk and boundary layer models.
The results are summarized in Sect.~\ref{sec:summary}.

\section{The data}
\label{sec:data}

In order to investigate the spectral variability of neutron star
LMXBs in the soft (high luminosity) state we have used the
RXTE/PCA \citep{rxte} observations of one of the bright Z-sources (Hasinger
\& van der Klis 1989) GX340+0. Our choice was defined by the requirement
that the PCA configuration  combined
sufficiently high energy resolution (large number of the energy channels) 
with good timing resolution and large total exposure time. 
We selected observations from proposal P20053 performed from Sep.21 to
Nov.4, 1997 with total exposure time of $\approx$178 ksec.  
The PCA configuration provided 31 energy channels in the 3--20 keV 
energy band with 2 msec time resolution. The detailed timing analysis
of these data was presented earlier by Jonker et al. (2000). 

The intensity-color diagram of GX340+0 is
presented in Fig.~\ref{fig:cid}. The Fig.~\ref{fig:pds_gx340} shows an example
of the power density spectrum, corresponding to the beginning of the
horizontal branch on the intensity-color diagram. The prominent QPO peak at
$\sim 25$ Hz corresponds to so called Horizontal Branch Oscillation (HBO).  

The kHz QPOs are known to be rather weak in the case of GX340+0
(especially the upper kHz QPO), and the available data on the source did not
have  enough sensitivity to study them in detail.
Therefore, to investigate the properties of two kHz QPOs and compare
them with the lower frequency variability we chose another bright LMXB,
4U1608-52. The data of proposal P30062 provide
energy and timing resolution adequate for this purpose.  Detailed
analysis of kHz QPOs in these observations was presented in
\cite{mendez99,mendez01}.  During these observations the source was in
the high luminosity state and had two strong kHz QPO peaks at
frequencies $\sim 600-800$ Hz and $\sim 900-1200$ and somewhat weaker
QPO at $\sim$45 Hz (Fig.~\ref{fig:pds_1608}).  Our aim is to compare
the spectral variability  corresponding to these three QPO
components.

The reduction of the PCA data was performed with the help of standard 
tasks from FTOOLS/LHEASOFT, version 5.1. The spectral approximations 
were done using XSPEC. In all spectral fits the interstellar
absorption was fixed at $NH=5\cdot 10^{22}$ cm$^{-2}$ and $NH=1\cdot
10^{22}$ cm$^{-2}$ for GX340+0 and 4U1608--52 respectively.

\begin{figure}
\includegraphics[width=0.5\textwidth]{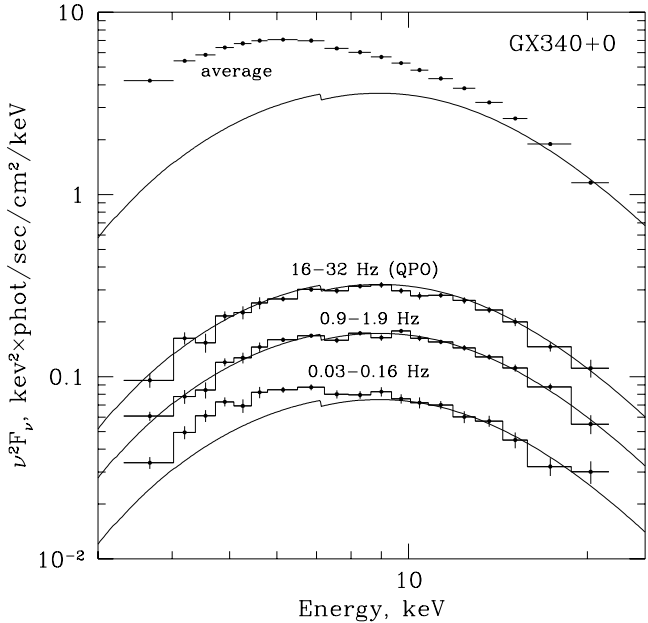}
\caption{Average and frequency resolved spectra of GX340+0 in the Horizontal
Branch (Z=0--1, see Fig.~\ref{fig:cid}). The solid lines show the
Comptonized emission spectrum rescaled to match the frequency resolved
spectra and the high  energy part of the average spectrum. The
parameters of the Comptonized spectrum are best fit to the frequency
resolved spectrum of the $\sim 25$ Hz QPO (Table \ref{tb:fit}).
\label{fig:freqres_gx340}}
\end{figure}

\begin{figure}
\includegraphics[width=0.5\textwidth]{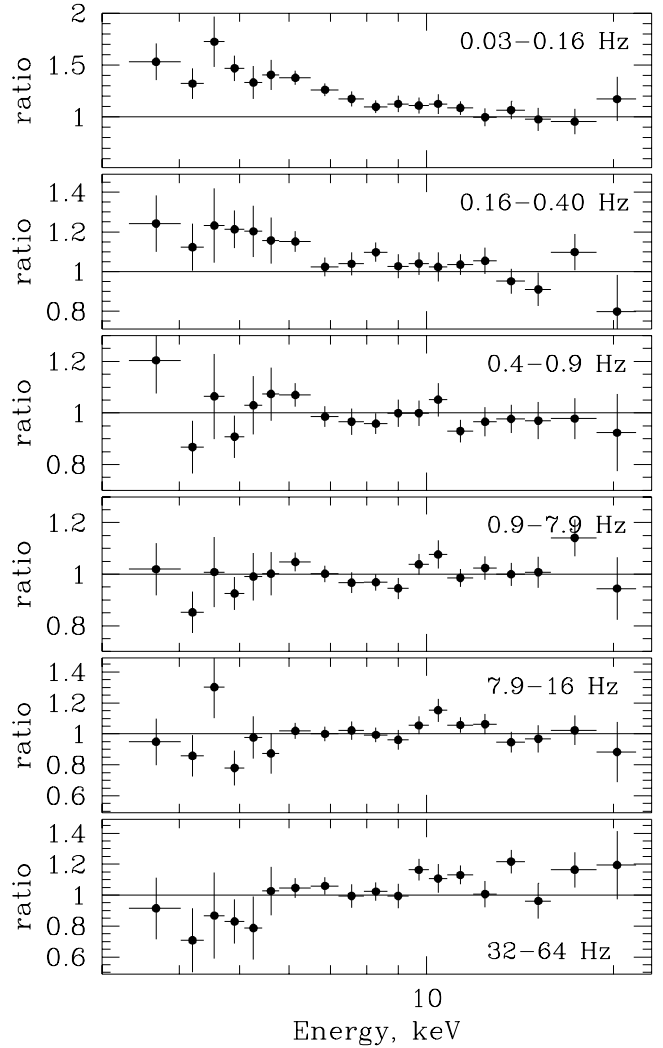}
\caption{GX340+0. Ratios of the frequency resolved spectra in various
Fourier frequency bands to that of the 16--32 Hz band, corresponding to the
Horizontal branch QPO. The data are from the Horizontal Branch of the
color--intensity diagram (Z=0--1), i.e. the same as in
Fig.~\ref{fig:freqres_gx340}. Note that the vertical scale is different
in different panels. 
\label{fig:freqres_gx340_ratios}}
\end{figure}

\section{Fourier frequency resolved spectra}
\label{sec:freqres_data}

\subsection{Low frequency continuum and QPO}

As is well known, the X-ray variability properties and parameters of
Z-sources depend significantly on the spectral state of the source
as given by its position on the Z-diagram, the highest level of
variability being observed at the beginning of the Horizontal Branch
(see e.g. \cite{jonker00}). Therefore throughout most of the paper we 
used the data averaged over Horizontal Branch of the
color-intensity diagram  (shown by a polygon in Fig.~\ref{fig:cid}).
The results for the normal branch are discussed in
Sect.~\ref{sec:discussion}. 
The power spectrum of GX340+0 on the horizontal branch is shown in
Fig.~\ref{fig:pds_gx340}.

The Fourier frequency resolved spectra 
in several frequency bands corresponding to the band limited continuum
noise component and  the $\sim 25$ Hz QPO are shown in
Fig.~\ref{fig:freqres_gx340} 
along with the conventional spectrum of the source averaged over the same
data. The ratios of the frequency resolved spectra to that of the QPO
are presented in Fig.~\ref{fig:freqres_gx340_ratios}. The spectra and
the ratios demonstrate that their shape depends on the
Fourier frequency at low frequencies and becomes independent of the
frequency at  $f\ga 0.5$ Hz.  
Similar behaviour was found by \citet{vdk86} in the case of another
bright LMXB -- Z-source GX5--1. In particular, it was noted, that
Horizontal Branch Oscillations and the low frequency noise have the
same, hard, fractional rms spectrum and that at lower frequencies
variabilty becomes softer.

The upper limits on the possible variations of the spectral shape with
Fourier frequency depend on the photon
energy and Fourier frequency and vary from $\sim 5\%$ to $\sim 25\%$
(Fig.~\ref{fig:freqres_gx340_ratios}).
Note that in the $0.5 \la f\la 30$ Hz frequency range these upper limits 
are determined by the statistical
uncertainties only. At higher frequencies, $f\ga 30$ Hz, there might be an
indication, although not statistically significant enough,  
of a frequency dependence of the spectral shape 
(lower panel in Fig.~\ref{fig:freqres_gx340_ratios}). Another conclusion from
the data presented in Fig.~\ref{fig:freqres_gx340}, important for the
following discussion,  is that  all frequency resolved spectra are
significantly harder than the average source spectrum.

\begin{figure}
\includegraphics[width=0.5\textwidth]{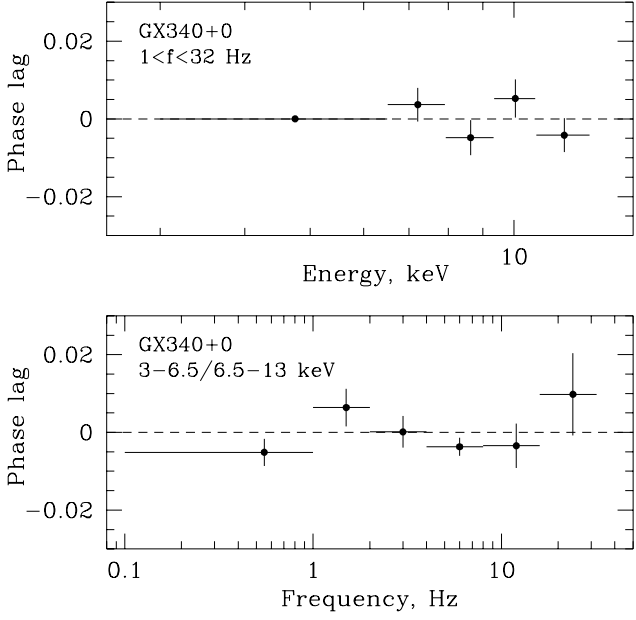}
\caption{Phase lags for GX340+0 in the Horizontal Branch of the
color--intensity diagram (Z=0--1) as function of energy ({\em upper panel})
and Fourier frequency ({\em lower panel}). The energy dependent phase lags
were computed in the 1--32 Hz frequency range, the frequency dependent lags
are between 3--6.5 keV and 6.5--13 keV energy bands. The phase is normalized
to 0--1 interval.
\label{fig:lags_gx340}}
\end{figure}

The phase lags as function of energy and Fourier frequency are shown in
Fig.~\ref{fig:lags_gx340}. No statistically significant phase lags were
detected in the Horizontal Branch of GX340+0 with upper limit of 
$\Delta\phi\sim 10^{-2}$, where phase $\phi$ is normalized to the interval
0--1 (as opposed to $0-2\pi$).

\begin{figure}
\vbox{
\includegraphics[width=0.5\textwidth]{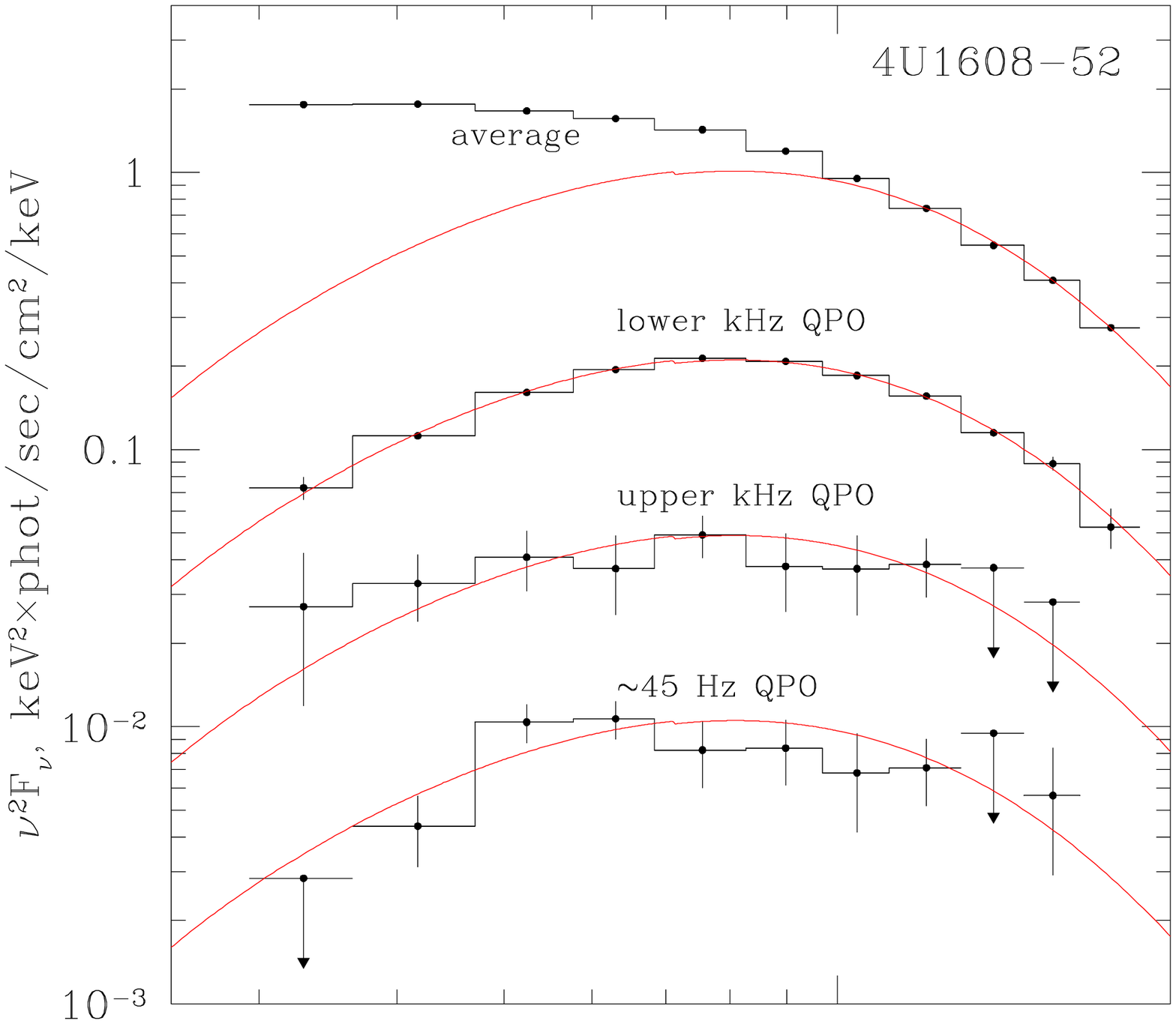}
\includegraphics[width=0.5\textwidth]{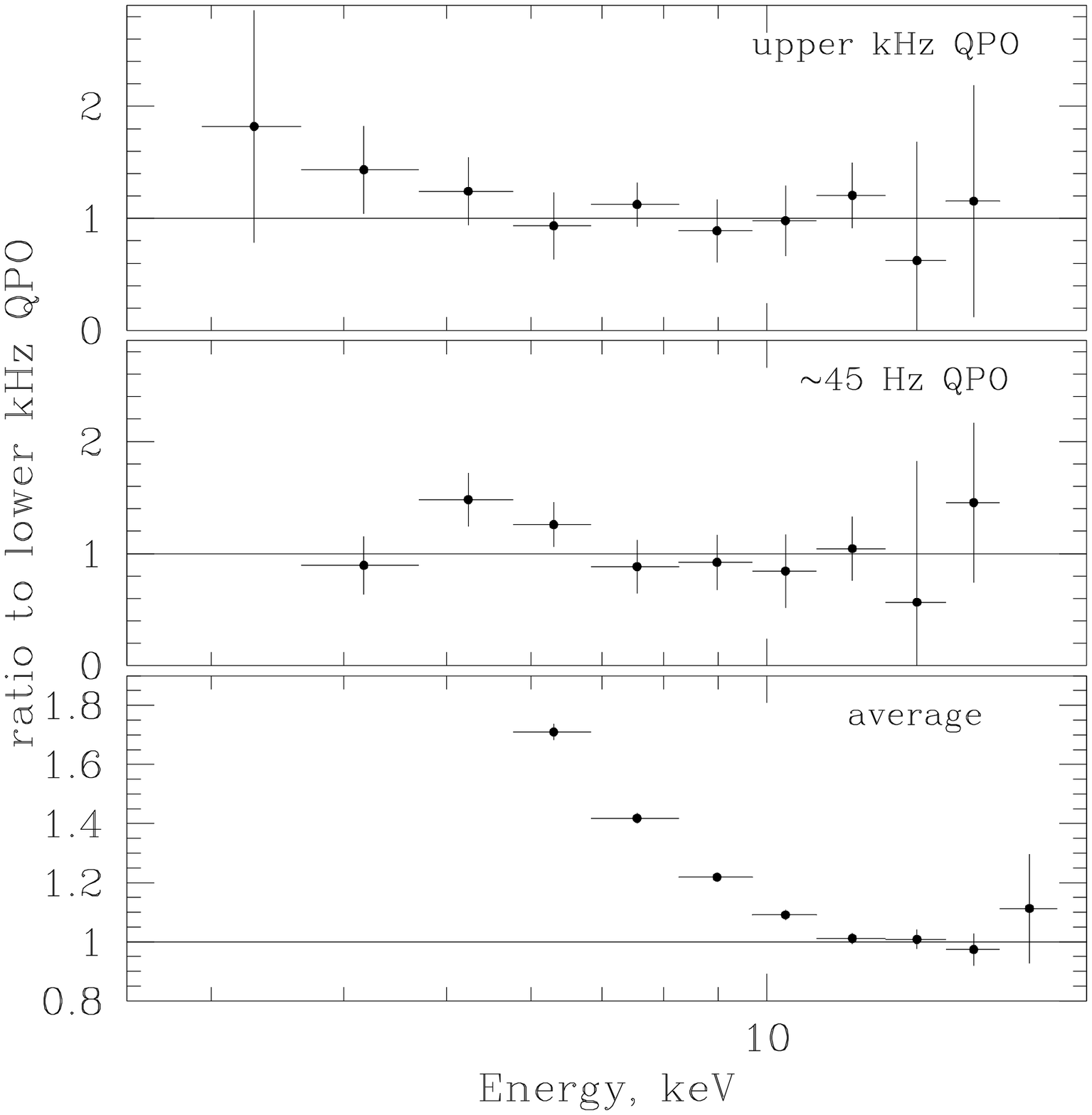}
}
\caption{{\em Upper panel:} Average and frequency resolved spectra
of 4U1608-52. The solid lines show  Comptonized disk spectra rescaled to
match the frequency resolved spectra and the high energy part of the
average spectrum.  
The parameters of the Comptonization spectrum are the best fits to the
frequency resolved spectrum of the lower kHz QPO (Table \ref{tb:fit}).
{\em Lower panel:} Ratios of the spectra shown in the upper panel
to the spectrum of lower kHz QPO.
\label{fig:freqres_1608}}
\end{figure}

\subsection{kHz QPOs}

As was mentioned above, the strength of the kHz QPOs in the power
spectrum of GX340+0 is insufficient to study their frequency resolved
spectra in detail. We therefore had to use for this purpose a different
source -- 4U1608-52 (Fig.~\ref{fig:pds_1608}).  As the frequency
of the kHz QPO varied during the RXTE observations we
used  ``shift-and-add'' technique \citep{mendez01} in order to improve
statistics. During the same observations the source showed 
a lower frequency QPO at $\sim$45 Hz as well. 

The Fourier frequency resolved spectra of the two kHz QPOs and the HBO along 
with their ratios are shown in Fig.~\ref{fig:freqres_1608}. As in the case
of GX340+0, all three  spectra are consistent with each other within
the statistical uncertainties. The upper 
limits on  possible variations of the spectral shape, however, are
significantly less constraining, $\sim 30-50\%$ at best.

\subsection{Summary of the observational results}
\label{sec:obs_sum}

The results of the above analysis  can be summarized 
as follows:  
\begin{enumerate}
\item The shape of the frequency resolved spectra does not depend or depends 
very weakly on the Fourier frequency. This includes the continuum band
limited noise component, the Horizontal Branch QPO and the kHz QPOs. The
constraints on  
possible Fourier frequency dependent variations of the spectral shape are
rather stringent for the band limited noise and Horizontal Branch
oscillations (GX340+0) -- $\la 5-25\%$ in the $0.5\ga f\la 30$ Hz frequency
range and significantly less restrictive for kHz QPO -- $\la 30-50\%$.
\item The Fourier frequency resolved spectra at all frequencies are
significantly harder than the average spectrum.
\item No statistically significant  phase lags were detected in the case of
GX340+0 with upper limits of $\Delta \phi \la 10^{-2}$ in the $0.1-30$ Hz
frequency range and $3-17$ keV energy range.
\item At low Fourier frequencies, $f\la 0.5$ Hz, frequency
dependent variations of the spectral shape become important -- the
frequency resolved spectra tend to become softer (GX340+0).
\end{enumerate}

\section{Interpretation of the Fourier frequency resolved spectra}
\label{sec:freqres_theory}

We show here that independence of the frequency
resolved spectra on the Fourier frequency and the smallness of the phase
lags observed in majority of bright LMXBs  (but see footnote
\ref{page:cygx2_fotnote})
require a particularly simple form of the spectral variability.

The constancy of the spectral shape with Fourier frequency implies that the
power spectrum $P(E,\omega)$ can be represented as a product of two
functions,  one of which depends on the energy and the other on the
frequency only. For convenience we write $P(E,\omega)$ in the form:  
\begin{equation} 
P(E,\omega)=S^2(E)\times f^2(\omega)
\label{eq:pds}
\end{equation} 
where non-negative functions $S(E)$ and $f(\omega)$ can be
directly determined from the frequency resolved spectra.
The Fourier image of the light curve $F(E,t)$ is:
\begin{equation} 
\hat F(E,\omega)=S(E)\times f(\omega)\times e^{i \phi(E,\omega)}
\end{equation}
In the general case the complex argument $\phi(E,\omega)$ can depend both
on Fourier frequency $\omega$ and energy $E$. 
If the phase lags between different energies are negligibly small,
$\phi$ depends on the Fourier frequency only and the 
Fourier image of $F(E,t)$ is: 
\begin{equation} 
\hat F(E,\omega)=S(E)\times f(\omega)\times e^{i \phi(\omega)}
\label{eq:ft}
\end{equation}
The light curve $F(E,t)$ can be computed via inverse Fourier
transform of $\hat F(E,\omega)$:
\begin{eqnarray}
F(E,t)=\int d\omega \hat F(E,\omega) e^{i\omega t}=\nonumber\\
=S(E)\times \int d\omega f(\omega) e^{i \phi(\omega)} e^{i\omega t}= 
\\
=S(E)\times f(t)\nonumber
\end{eqnarray}
An arbitrary function of energy can obviously be added to the above
expression. Thus, the light curve at energy $E$ should satisfy the 
equation:
\begin{equation} 
F(E,t)=S_0(E)+S(E)\times f(t)
\label{eq:lc}
\end{equation}
i.e. the light curves at different energies are related by a
linear transformation. 

Eq.~(\ref{eq:lc}) significantly restricts the pattern of the spectral
variability. 
Suppose that the spectrum of the source at any given moment of time
can be represented as:
\begin{equation} 
F(E,t)=S_0(E)+A(t)\times S(E,p(t))
\end{equation}
where $A(t)$ is varying normalization, and $p(t)$ represents
variations of the spectral parameter, on which the spectral flux at a given
energy depends non-linearly (e.g. temperature, Thompson optical depth
etc.). Using Taylor expansion:
\begin{eqnarray}
F(E,t)=S_0(E)+A(t)\times S(E,p(t))=\hfill \nonumber  \\
= S_0(E)+A(t)\times S(E,p_0)+ \nonumber \\
+ A(t) \times \frac{\partial S(E,p_0)}{\partial p}\times (p(t)-p_0) + 
\label{eq:taylor}
\\
+ A(t) \times \frac{1}{2} \frac{\partial^2 S(E,p_0)}
{\partial p^2}\times (p(t)-p_0)^2+... \nonumber
\end{eqnarray}
In order to satisfy Eq.~(\ref{eq:lc}), one of the following two
conditions should be fulfilled: 
(i) $p(t)=p_0={\rm const}$, i.e. only the normalization varies
with time, or  (ii) the normalization is constant and variations of
the spectral 
parameter are sufficiently small, so that the terms higher than
linear in the Taylor expansion can be neglected. 
Simultaneous variations of comparable amplitude  of the normalization
and the spectral parameter (or of any two spectral parameters)  would
be consistent with observations only in the case of $A(t)\times
(p(t)-p_0)={\rm const}$ or in the case of specific shape of the power
density spectrum of $A(t)$ (e.g. flat).

\begin{figure}
\includegraphics[width=0.5\textwidth]{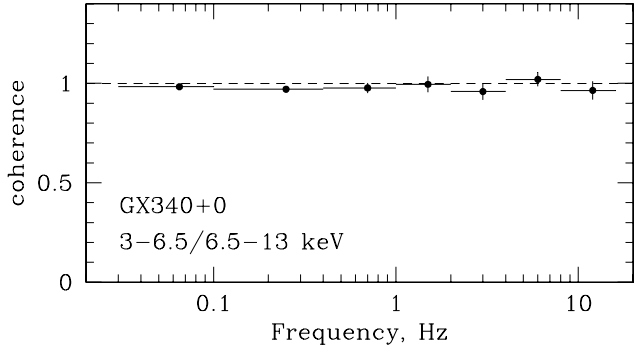}
\caption{GX340+0: Coherence between the light curves in the 3--6.5
  and 6.5--13 keV energy bands as function of frequency. No correction
  for the dead time effects has been made.  
\label{fig:coh_gx340}}
\end{figure}

From Eq.~(\ref{eq:lc})  it  follows that the coherence
of the signals in any two energy bands is exactly unity, as
they are related by a linear transformation. This is in agreement
with the observed behaviour -- as illustrated by an example of
GX340+0, shown in Fig.~\ref{fig:coh_gx340}, the  coherence between the
light curves in  3--6.5 and 6.5--13 keV energy bands  is close to unity
(see also \citealt{vaughan94,  vaughan99, dieters00} for results for
other luminous LMXBs). 

\subsection{Case of black hole binaries}

The spectral variability 
in the black hole binaries is more complicated than described by
Eq.~(\ref{eq:lc})  because of the strong dependence of the shape of their
frequency resolved spectra upon the Fourier frequency
\citep{freq_res99},\footnote 
{
With a possible exception of the soft state, e.g. in the soft state of Cyg
X-1 \citep{chur01} found that the shape of the power density spectra does
not depend on energy and that the spectral variability satisfies the
Eq.~(\ref{eq:lc}). 
} 
rather than non-zero
time/phase lags observed. Indeed, typical values of the phase lags
measured for Cyg X-1 are negligible in this context, $\Delta\phi\la
{\rm few}\times 10^{-2}$ \citep*[e.g][]{kotov}. Therefore 
\[
\phi(E,\omega)=\phi_0(\omega)+\phi_1(E,\omega)
\]
where
\[
\phi_1(E,\omega)<<\phi_0(\omega)
\]
and Eq.~(\ref{eq:ft}) and therefore Eq.~(\ref{eq:lc}) will hold with
sufficient accuracy if condition of Eq.~(\ref{eq:pds}) is fulfilled.

\section{Application to bright LMXBs}
\label{sec:freqres_applications}

\subsection{Frequency resolved spectra and nature of the
variable component} 
\label{sec:freqres_applications1}

Thus, observed properties of the luminous LMXBs
(Sect.~\ref{sec:obs_sum}) require a particularly simple form of the
spectral variability on time scales from $\sim$ sec to $\sim$ msec,
described by Eq.~(\ref{eq:lc}). 
The term $S_0(E)$ in Eq.~(\ref{eq:lc})  represents constant
(non-variable) part of the source 
spectrum and $f(t)$ represents either (i) variations of the
normalization of the variable spectral component
\citep[e.g.][]{chur01} or (ii) small variations of the spectral
parameter \citep[e.g.][]{mendez0614}.

If variability of the X-ray flux is caused mainly by 
variations of the normalization, it follows from
Eq.~(\ref{eq:pds}), that the spectrum of the
variable component $S(E)$ is identical 
in shape to the frequency resolved spectrum, i.e. can be directly
determined from observations.
The spectrum   of the non-variable component can, in principle, be
obtained  subtracting  $S(E)$ from the averaged spectrum with appropriate 
renormalization.  

In the  second case (small variations of the spectral parameter), the
energy dependence of the frequency resolved spectrum 
$S(E)$ corresponds to the first derivative of the spectrum with
respect to the parameter $p$,  $\frac{\partial S(E,p)}{\partial p}$,
which might   
differ significantly from the spectrum itself. Indeed, considering
a Wien spectrum, $S(E,T)=E^2 e^{-E/T}$, as an example
(cf. \citet{mendez0614}), the intensity variations are: 
\begin{eqnarray}
F(E,t)=S(E,T_0)+\frac{\partial S(E,T_0))}{\partial T}\times
\delta T(t)+...= 
\label{eq:wien}
\\
= S_0(E)+S_0(E)\times\frac{E}{T_0}\times \frac{\delta T(t)}{T_0}+...
\nonumber
\end{eqnarray}
Therefore in Eq.~(\ref{eq:lc}) $S(E)=S_0(E)\times E/T_0$,
i.e. is harder than the average spectrum.
It is intuitively obvious, however, that in order the linear expansion
to be valid, the variations amplitude at any energy should be smaller
than the average flux.
Indeed, for the quadratic term in
Eq.~(\ref{eq:taylor}) to be neglectable, the following condition should
be satisfied:  
\begin{eqnarray}
\frac{E}{T_0}\times \frac{\delta T(t)}{T_0} \ll 1
\end{eqnarray}
i.e. (cf. Eq.~(\ref{eq:wien})) the frequency resolved spectrum in any
frequency range and at any 
energy should be small in comparison with the average spectrum of the
variable component. Equivalently, fractional rms of flux variations
computed with respect to the average flux of the variable component
(as opposite to the total average flux) should be less than 1.

In the first  case identification of the
variable component would be easy and unambiguous. 
However, no distinction between the two possibilities outlined above
can be made based solely on the results of the frequency resolved
analysis, and additional considerations should be taken into account.

\begin{figure*}
\hbox{
\includegraphics[width=0.5\textwidth]{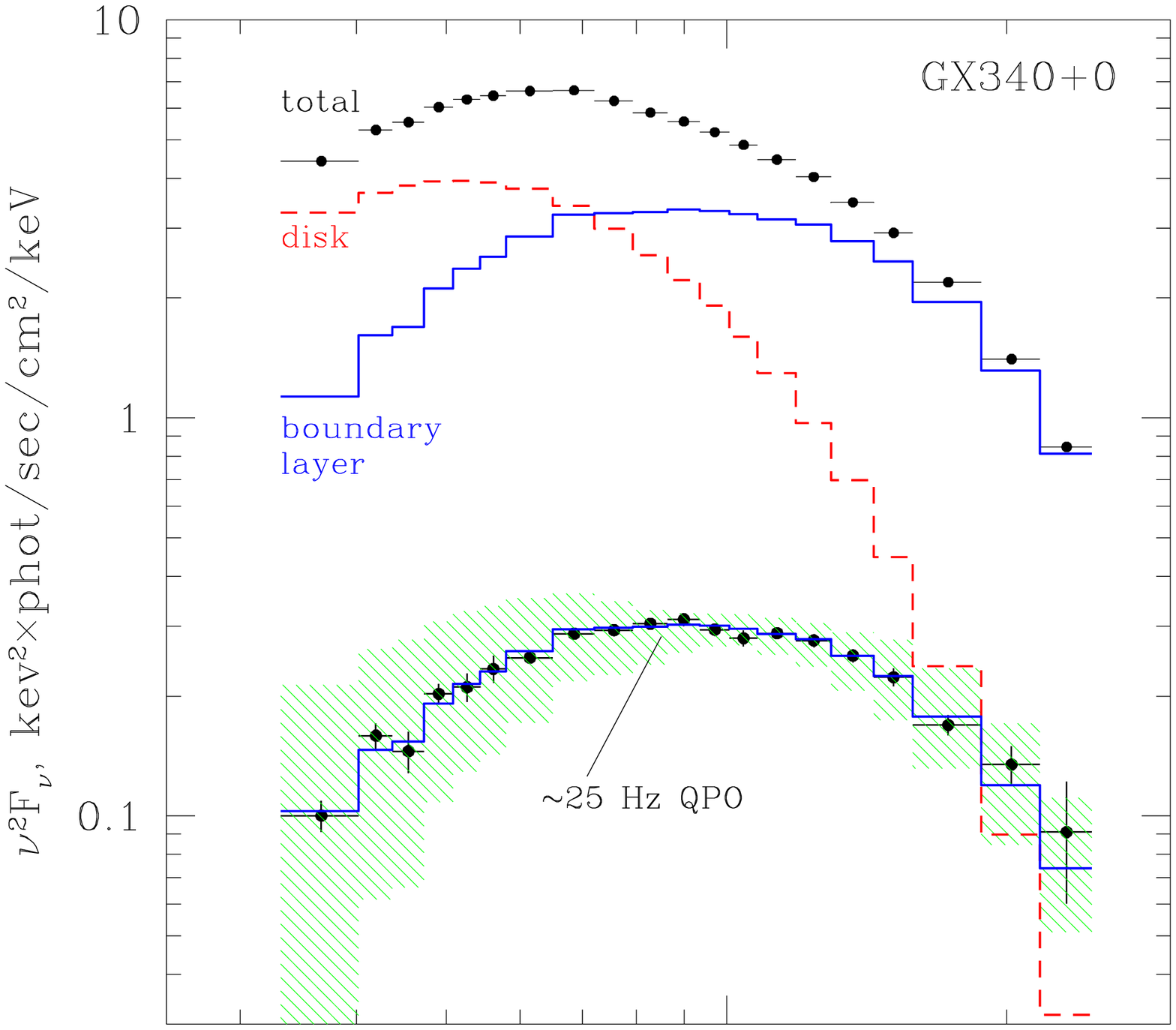}
\includegraphics[width=0.5\textwidth]{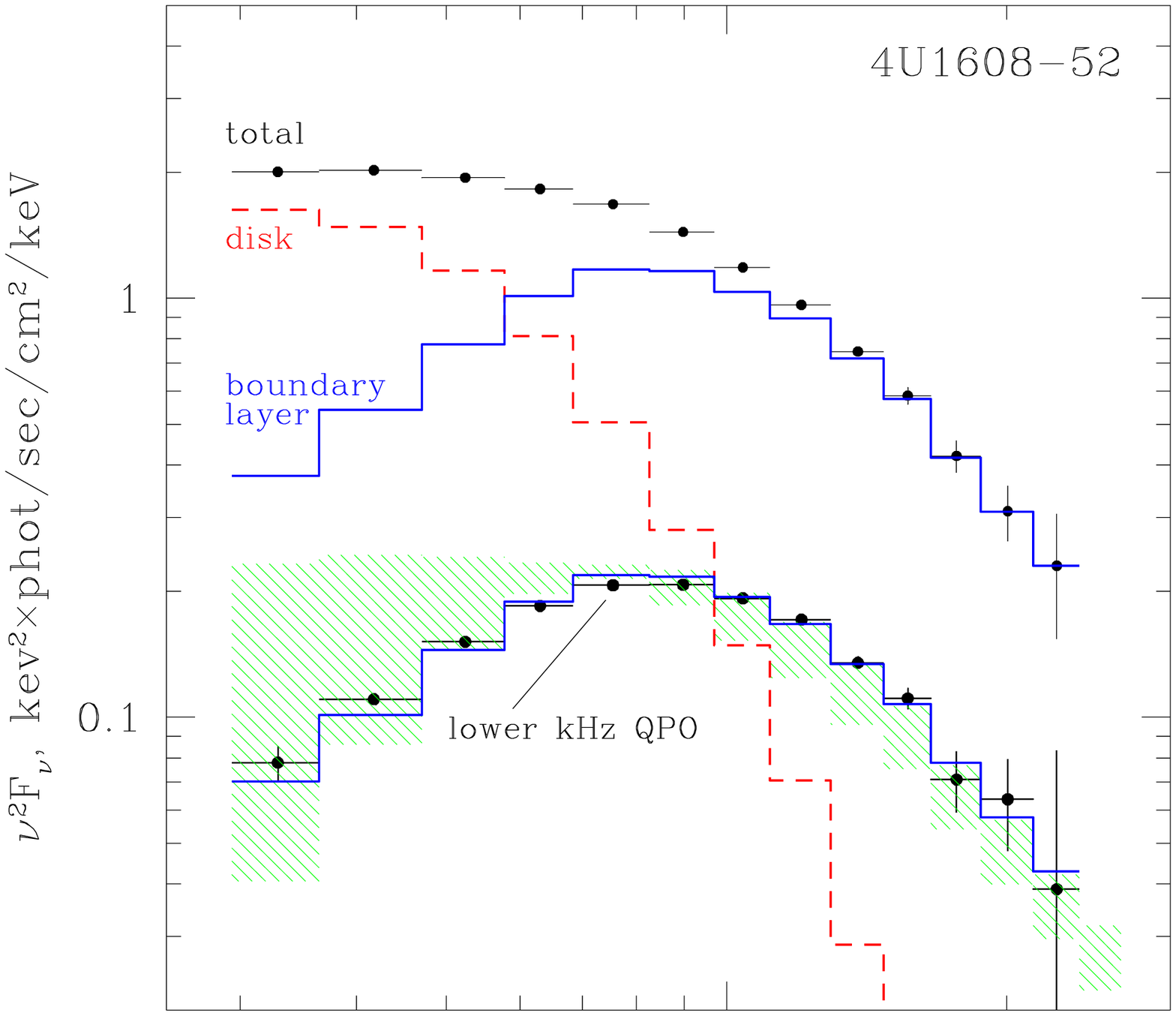}

}
\hbox{
\includegraphics[width=0.5\textwidth]{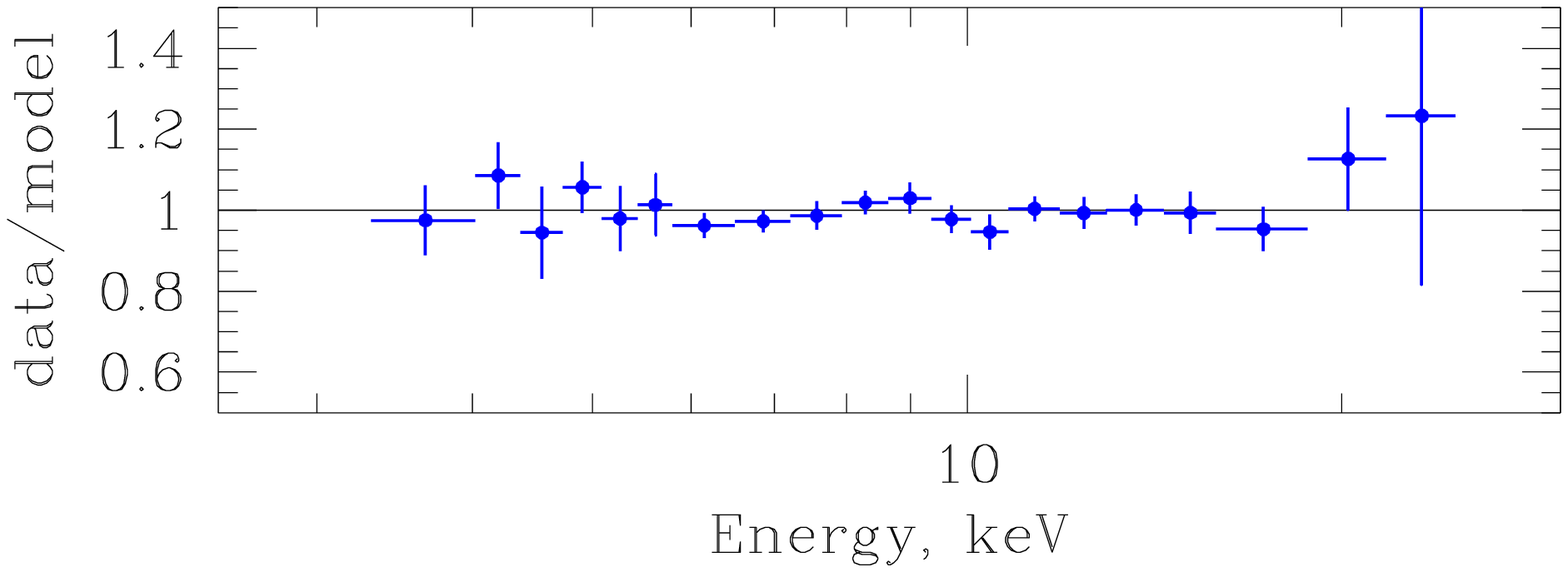}
\includegraphics[width=0.5\textwidth]{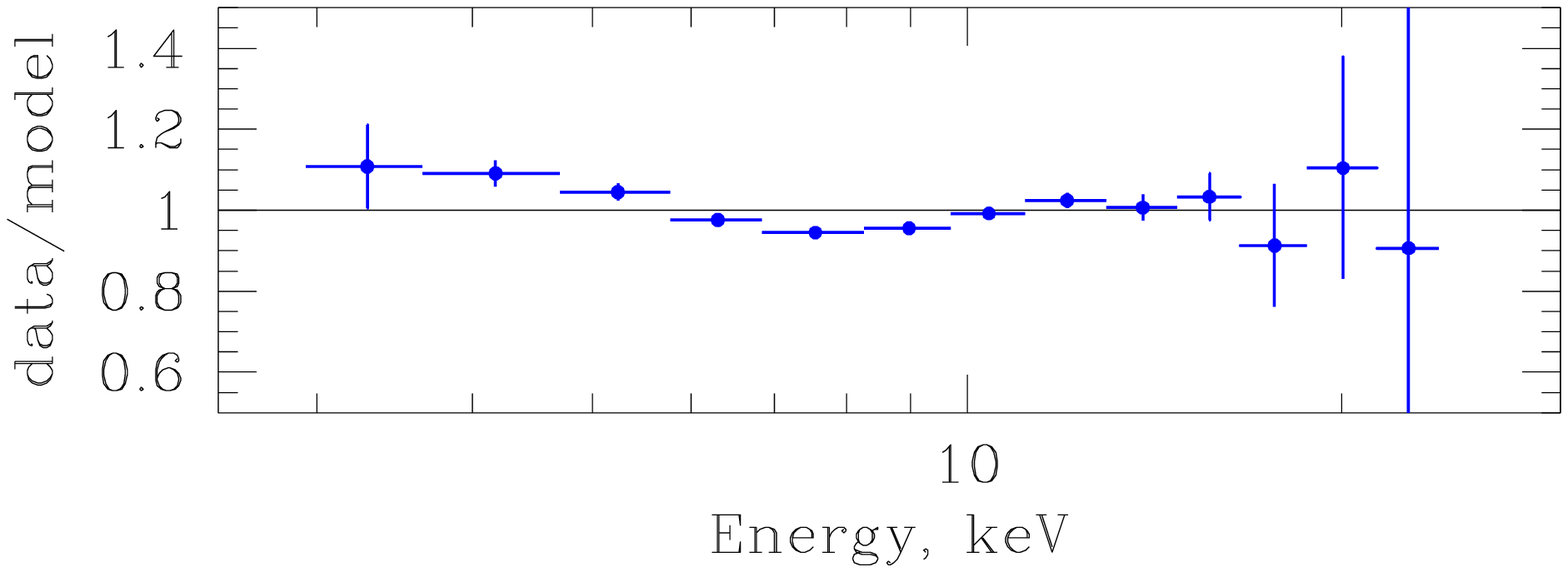}
}
\caption{The average and frequency resolved spectra of  GX340+0 ({\em
left}) and 4U1608--52 ({\em right}). The GX340+0 data are from
horizontal branch of the color--intensity diagram, i.e. the same as
used before. The shaded area shows a plausible range of the boundary
layer spectral shape calculated subtracting the predicted disk
spectrum from the total spectrum and renormalizing the residual to the
total energy flux of the frequency resolved spectrum (see
Sect.~\ref{sec:disk_bl} and Table \ref{tb:disk_range} for
details). The dashed  histogram shows the accretion disk  spectrum
with parameters from Table \ref{tb:fit} (Sect.~\ref{sec:fit}), the
upper solid histogram shows the difference between the total and
accretion disk spectrum ($\approx$ boundary layer spectrum). 
The lower solid histogram is the same but scaled to the total energy flux
of the frequency resolved spectrum. The lower panels show ratio of the
frequency resolved spectrum to the lower histogram.
\label{fig:blspe}}
\end{figure*}

\subsection{Emission from the boundary layer and accretion disk}
\label{sec:disk_bl}

Based on theoretical grounds \citep{ss86, kluzniak, inogamov99,
popham01} and observational results \citep[e.g.][]{mitsuda84,
mitsuda86, gierlinski00} it is expected that in the case of
accretion onto  a slowly rotating weakly magnetized neutron 
star at sufficiently high mass accretion rates there are two major
components of the accretion flow:  
\begin{enumerate}
\item the optically thick accretion disk extending close to the
surface of the neutron star or last marginally stable orbit
\item the boundary layer \citep[e.g.][]{popham01} or spreading layer
\citep{inogamov99}  near the surface of the neutron star. In this
layer the accreting matter decelerates to the spin frequency of the
star and spreads over its surface.  
\end{enumerate}
Simple theoretical arguments, taking into account energy released in
the accretion disk and in the boundary layer, and difference in their 
emitting areas  suggest that the spectrum of the boundary layer
emission should be noticeably harder than that of the optically
thick accretion disk  \citep[e.g.][]{mitsuda84,greb}.   
Considering the restrictions on the character of spectral variability
imposed by Eq.~(\ref{eq:lc}), it is natural to assume, that the
observed variations of the X-ray flux should originate in one of these
components. Furthermore, X-ray 
variability should be mainly caused by either variations of the total
luminosity with nearly constant spectral shape or by small
variations of the spectral shape.
In order to distinguish between these possibilities and to identify
the nature of the varying component we consider 
below theoretical expectations for the disk and boundary layer spectra
and compare them with the observed frequency resolved spectra.

Due to complexity of the boundary/spreading layer problem, no
models capable to directly predict its emission spectrum 
exist yet. 
An attempt of solve the radiation transfer problem using density
and temperature profiles obtained in the hydrodynamical simulations  
\citep[e.g.][]{popham01} failed to reproduce, even
qualitatively, the observed X-ray spectra and demonstrated importance
of the self-consistent treatment \citep{greb}. 

Significantly better progress has been achieved with the spectra of
accretion disks  \citep{ss73, shimura_takahara95,
ross96}. It has been shown that at sufficiently high mass accretion
rates the effects of Compton scattering can be approximately accounted
for by introducing a dilution factor to describe deviation of the spectra
emitted at each radius of the disk from blackbody
\citep{shimura_takahara95, ross96}. Additional modifications of the
disk spectrum arise due to gravitational redshift and Doppler effects 
\citep{grad, ross96, diskns}. Relatively simple models of multicolor
disk type which account for the effects of Compton scattering with a
simple color-to-effective temperature ratio  turned out to be
successful in describing the accretion disk spectra observed in the
high state of black hole systems \citep[e.g.][]{grad, gierlinski97}.

\begin{table}
\caption{Range of parameters used to estimate  plausible range of
the accretion disk and boundary layer spectra
(Fig.~\ref{fig:blspe})}
\label{tb:disk_range} 
\begin{centering}
\begin{tabular}{lcc}
\hline
Parameter & GX340+0 & 4U1608-52\\
\hline
D, kpc		& 8.5--10.5	& 3.5--4.5 		\\
$F_{\rm 0.1-30~keV}$ $^1$
& $2.7\cdot 10^{-8}$ & $8.8\cdot10^{-9}$\\
$\cos(i)$	& \multicolumn{2}{c}{0.3--0.7}	\\
$T_{\rm col}/T_{\rm eff}$ & \multicolumn{2}{c}{1.6--2.0}	\\
$\nu_{\rm NS}, Hz$	& \multicolumn{2}{c}{0--700}	\\
\hline
$\eta_{\rm tot}$ $^2$	& \multicolumn{2}{c}{0.213--0.116}\\
$f_{\rm disk}$ $^2$	& \multicolumn{2}{c}{0.26--0.57}\\
$f_{\rm BL}$ $^2$	& \multicolumn{2}{c}{0.74--0.43}\\
\hline
\end{tabular}\\
\end{centering}
1 -- absorption corrected, erg/s/cm$^2$ \\
2 -- derived values of the total accretion efficiency and fraction of
the energy released in the disk and boundary layer respectively,
corresponding to the considered  range of the neutron star spin
frequency, for a $1.4 M_{\sun}$ neutron star, EOS FPS \citep{sibg00}. \\
\end{table}

Therefore we use the predicted disk spectrum as a starting point and
subtract it from the total spectrum in order to calculate expected
spectrum of the boundary layer. To estimate the plausible range of the
boundary layer spectra we investigate the parameter space of the
accretion disk model.
This approach enables one to predict the
disk and the boundary layer spectra based on the  
observed X-ray flux and spectrum and
very generic system parameters, such as neutron star spin frequency or
the source distance. 
For this analysis we adopt the general relativistic accretion disk
model by \citet{grad} (the ``grad model'' in XSPEC). 
The parameters of the model are: the source distance $D$,
mass of the central object $M_{\rm NS}$, disk inclination 
angle $i$, the mass accretion 
rate $\dot{M}$ and the color-to-effective temperature ratio 
$f=T_{\rm col}/T_{\rm eff}$. 
The source distance was varied in the range 8.5--10.5 kpc (GX340+0)
and 3.5-4.5 kpc (4U1608--52), the disk inclination -- in the range of
$\cos i=0.3-0.7$,  color-to-effective temperature ratio -- in the
range 1.6-2.0 \citep{shimura_takahara95}.

In order to estimate the plausible range of  mass accretion rates
we used the total accretion efficiency $\eta_{\rm tot}$ 
and relative energy release fractions of the disk, $f_{\rm disk}$, and
the neutron star surface, $f_{\rm BL}$, calculated by \citet{sibg00}.  
These account exactly for the space-time metric around a neutron star
with a given spin frequency and equation of state.
In particular we used their results for a neutron star
with gravitational mass of 1.4 $M_{\sun}$ described by FPS equation of
state (their eqs. (1) and (2)). In computing the relation between the
observed luminosity and the mass accretion rate we ignored the light
bending and abberation effects, the real geometry of the spreading
layer and shadow cast by the neutron star on the disk and boundary
layer: 
\begin{eqnarray} 
L_{\rm obs}=\left(d(i) f_{\rm disk}+
d(\frac{\pi}{2}-i) f_{\rm BL}\right)\times \eta_{\rm tot} \dot{M}c^2
\label{eq:mdot}
\end{eqnarray} 
where emission diagram is described by Chandrasekhar--Sobolev
law $d(i)\approx 3/7 (1+2.06 \cos i) \cos i$. 
The luminosity $L_{\rm obs}$ was computed from the
absorption corrected fluxes extrapolated to 0.1--30 keV energy
range. The total correction factor from observed 3--20 keV flux was
$\sim 2$ in both cases. Therefore, unless a new
spectral component appears outside the 3--20 keV PCA energy range,
the uncertainty of the flux correction should not exceed few tens of 
per cent. The accretion efficiency $\eta_{\rm tot}$ and the disk and
boundary layer fractions $f_{\rm disk}$ and $f_{\rm BL}$ depend,
obviously, on the neutron star spin, which was varied in the range
0--700 Hz assuming that the neutron star and disk co-rotate.

The investigated range of the disk parameters is summarized in Table
\ref{tb:disk_range}.  Each predicted disk spectrum was subtracted from
the total spectrum and the residual ($\approx$boundary layer spectrum)
was normalized to the observed 3--20 keV energy flux of 
the frequency resolved spectrum.  
The resulting range of the boundary layer spectra for two sources is
shown in Fig.~\ref{fig:blspe} as the shaded area along with the average
and frequency resolved spectra of the 25 Hz and kHz QPO and the best
fit 
model described in Subsect.~\ref{sec:fit}.  The lower panels in
Fig.~\ref{fig:blspe} show ratio of the frequency resolved spectrum to
predicted boundary layer spectrum.
Different models of the disk emission, e.g. ignoring the relativistic
effects (the ``diskpn'' model in XSPEC, \citealt{gierlinski97}) or model of
\citet{ross96} which  more accurately accounts for the Compton scattering
effects, but assuming Newtonian dynamics,  change the details but
do not alter the general conclusion. 
The close similarity of the expected boundary layer and the
observed frequency resolved spectra strongly suggests that 
X-ray variability is caused by variations of the luminosity of the
boundary layer component.  It also implies that the spectrum of the
boundary layer emission does not change significantly in the course of
these variations (cf. example given by Eq.~(\ref{eq:wien})).

\begin{figure}
\includegraphics[width=0.5\textwidth, clip]{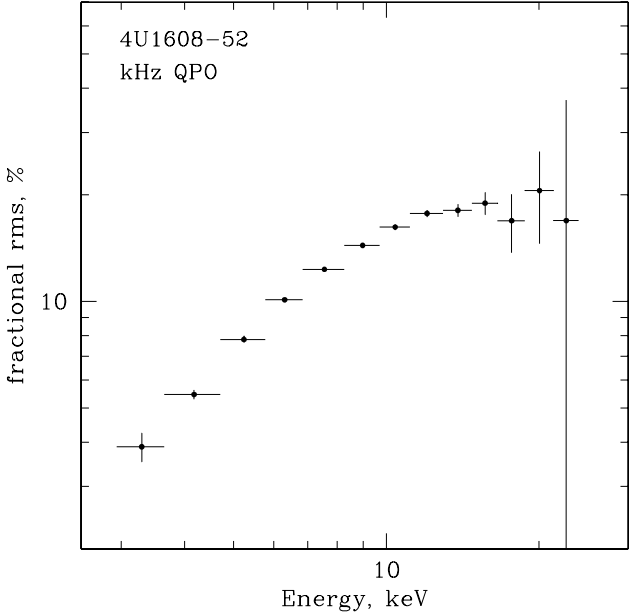}
\caption{The energy dependence of the fractional rms of the kHz QPO in
4U1608--52. Flattening above $\sim 10$ keV indicates, that the energy
spectrum of the oscillations at these energies has the same shape as
the total spectrum. 
\label{fig:rms_1608}}
\end{figure}

On the other hand it is obvious from Fig.~\ref{fig:blspe} that the
disk spectrum is significantly softer and can not reproduce the
observed frequency resolved spectrum. Qualitatively, this fact does
not rule out the possibility that X-ray variability
is due to variations of the disk emission spectrum.
Indeed,  variations of e.g. disk temperature
can result in the frequency resolved spectra significantly different
(in particular, harder, cf. Eq.~(\ref{eq:wien})) than the average
spectrum. However, as discussed in
Sect.~\ref{sec:freqres_applications1}, in order   
to maintain constancy of the shape of the frequency resolved spectra
and zero time lags between different energy channels it is required,
that  frequency resolved spectra in any frequency range are $\ll$ than
the average spectrum of the variable component.
This condition is obviously violated at $E\ga 10-15$ keV
(Fig.~\ref{fig:blspe} -- cf. the disk  and the frequency
resolved spectra).

Considering the quick decline of the disk emission above $\sim 10$ keV,
another test of the suggested interpretation might be the
behavior of the frequency resolved spectrum (or fractional rms) at
high energy, $\ga 10-15$ keV. Indeed, as at $E \ga 10-15$ keV the
total spectrum is dominated by the boundary layer emission,
the fractional rms in the above scenario  is expected to become
independent of the photon energy. 
Due to insufficient energy coverage this can not be checked
in the case of GX340+0 where the disk temperature is rather high. 
Flattening of the fractional rms is clearly seen in 4U1608--52 (lower-most
panel in Fig.~\ref{fig:freqres_1608} and Fig.~\ref{fig:rms_1608}),
having $\sim 10$  times lower mass accretion rate and correspondingly
lower disk temperature and softer disk spectrum.  Similar behavior
was observed by \citet{mitsuda84} for Sco X-1 using TENMA data.

Thus, we conclude that the bulk of the X-ray variability observed
in GX340+0 and 4U1608--52 and, presumably, in other luminous LMXBs, on
$\sim$ second -- millisecond time scales is due to variations of
the boundary layer luminosity. The shape of the boundary layer
emission spectrum remains nearly constant in the course of these
variations. Therefore the energy spectrum of the variable component -- the 
frequency resolved spectrum should be representative, to some
accuracy, of the spectrum of the boundary layer emission.  This can be
used to separate the boundary layer and the accretion disk
contribution to the total spectrum and permits to check quantitatively
the predictions the accretion disk and boundary layer models. 
It also opens the  possibility to measure relative contributions of
these two components of the accretion flow to the total observed X-ray 
emission.

\begin{table}
\caption{Parameters of the spectral modeling}
\label{tb:fit} 
\begin{tabular}{lcc}
\hline
\hline
Parameter & GX340+0 & 4U1608-52\\
\hline
D, kpc		& 8.5			& 4.0 				\\
NH, cm$^{-2}$	& $5\cdot 10^{22}$ 	& $1\cdot 10^{22}$		\\
$F_{\rm3-20~keV}$ $^1$ erg/s/cm$^2$ & $1.4\cdot 10^{-8}$ & $4.2\cdot 10^{-9}$\\
$F_{\rm 0.1-30~keV}$ $^2$ erg/s/cm$^2$ & $2.7\cdot 10^{-8}$ & $8.8\cdot10^{-9}$\\
$L_{\rm 0.1-30 keV}$ $^2$ erg/s & $2.3\cdot 10^{38}$	& $1.7\cdot 10^{37}$\\
$\dot{M}$ $^3$,  g/s	& $3.1\cdot 10^{18}$	& $2.0\cdot 10^{17}$	\\
\hline
\hline
\multicolumn{3}{c}{QPO frequency resolved spectra ($\approx$boundary layer)$^4$}\\
\hline
\multicolumn{3}{c}{power law with exponential cutoff (phabs$\times$cutoffpl)}\\
\hline
$\alpha$	&$-0.55\pm0.16$	&$-1.28\pm 0.13$\\
$E_{\rm f}$, keV	&$3.3\pm0.2$	&$2.4\pm 0.1$	\\
$\chi^2$/dof	& 13.9/16	& 4.8/9		\\
\hline
\multicolumn{3}{c}{Comptonization model (phabs$\times$comptt)}\\
\hline
$kT_{\rm bb}$, keV	& $1.3\pm0.2$	& $1.4\pm0.4$	\\
$kT_{\rm e}$, keV	& $3.1_{-0.3}^{+0.9}$	& $2.6_{-0.3}^{+\infty}$\\
$\tau$		& $6.0_{-2.1}^{+1.8}$	& $6.7_{-5.3}^{+5.7}$	\\
$\chi^2$/dof	& 11.3/15	& 4.0/8		\\
\hline
\hline
\multicolumn{3}{c}{average spectra$^4$}\\
\hline
\multicolumn{3}{c}{phabs$\times$(grad+pexrav+gaussian)}\\
\hline
inclination$^5$	& $60\degr$	&	$70\degr$\\
$T_{\rm col}/T_{\rm eff}$ $^5$ & 1.7	& 1.8	\\
$\dot{M}$, $10^{18}$ g/s	& $3.0\pm 0.04$	& $0.34\pm 0.01$\\
$\dot{M}/\dot{M}_{\rm Edd}$ $^6$	& $\approx 0.9$	& $\approx 0.1$\\
$\Omega/2\pi$	& $0.27\pm0.07$		& $\la 0.1$\\
$EW$ $^7$, eV	& $48\pm 10$ 	& $147\pm 17$\\
$L_{\rm BL}/L_{tot}$, 3--20 keV	&	47\%	&	57\%	\\
\hline
\hline
\end{tabular}\\
\\
1 -- observed; 2 -- absorption corrected;
3 -- calculated from the total unabsorbed luminosity 
using the accretion efficiency for the neutron star spin frequency of
$\nu_{\rm NS}=500$ Hz  \citep[][see Sect.~\ref{sec:disk_bl},
Eq.~(\ref{eq:mdot})]{sibg00};  
4 -- the details of spectral modeling are given in Sect.~\ref{sec:fit};
5 -- fixed at fudicial value;
6 -- assuming $\dot{M}_{\rm Edd}=2\cdot 10^{38}/c^2\eta_{\rm disk}
\approx 3.5\cdot 10^{18}$ g/s, where  $\eta_{\rm disk}=0.066$ --
accretion disk efficiency for $\nu_{NS}=500$ Hz and
$M_{NS}=1.4M_{\sun}$; 
7 -- line energy and width were fixed at $E_{\rm line}=6.7$ keV and
$\sigma_{\rm line}=0.5$ keV
\end{table}

\subsection{Spectral modeling} 
\label{sec:fit}

As small variations of the spectral shape of the boundary layer
emission in the course of flux variations can not be excluded and are
likely to take place, the frequency resolved spectra represent 
the boundary layer emission with a certain accuracy only.
This should be kept in mind while interpreting the spectral fitting
results described in this  subsection. 
However, the reasonable quality of the spectral fits and values
of the spectral parameters for two sources with significantly
different mass accretion rate provide additional indirect support to
the main conclusion of this section.

Below we approximate the frequency resolved spectrum ($\approx$
spectrum of the boundary layer emission) with a plausible model and
then fit the total spectrum adding the accretion disk component. This
procedure is equivalent to subtracting the renormalized frequency resolved
spectrum from the total spectrum, with the renormalization coefficient 
determined by requirement to minimize $\chi^2$.

\begin{figure}
\centering{
\vbox{
\includegraphics[width=0.42\textwidth, clip]{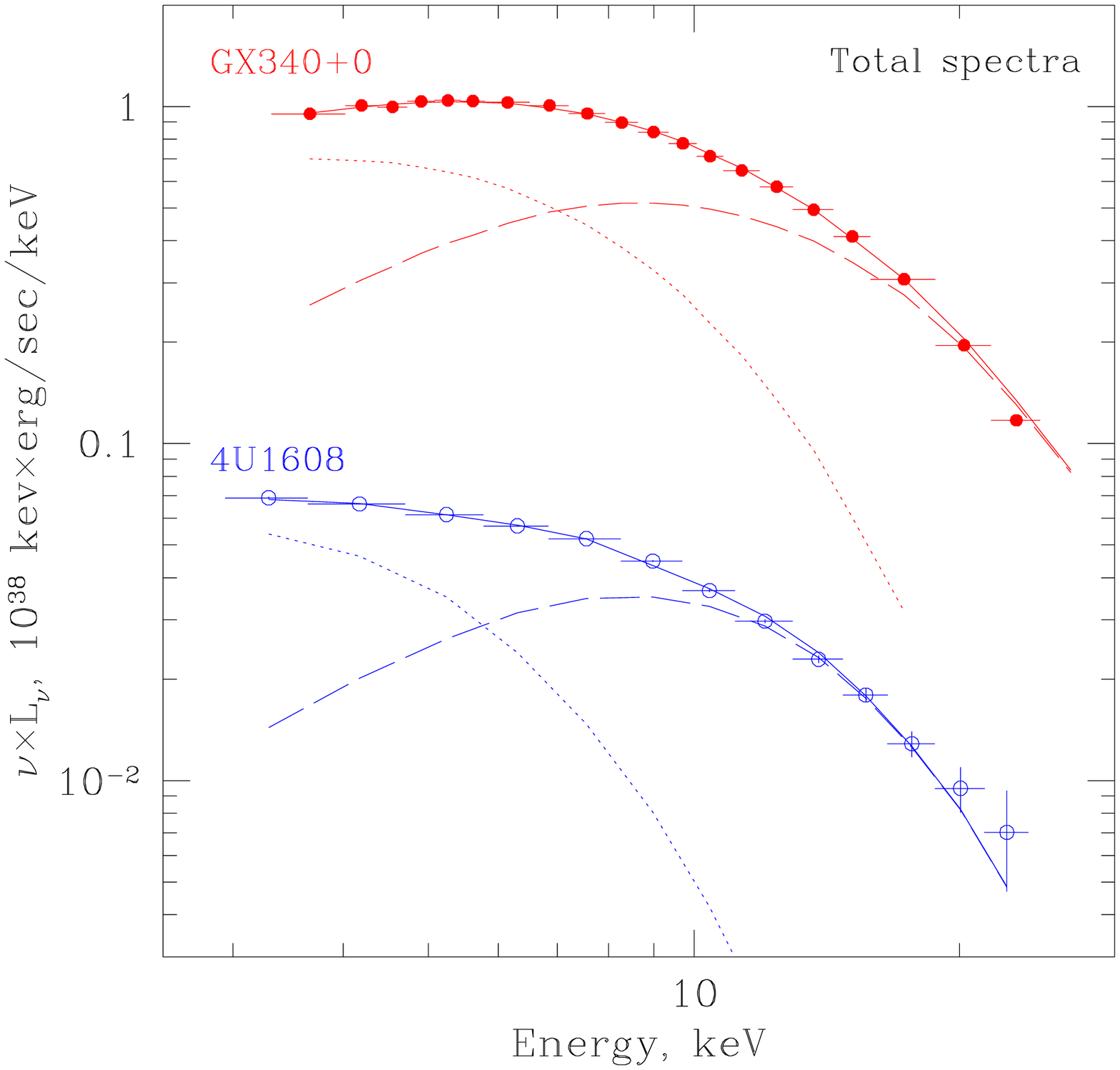}
\includegraphics[width=0.42\textwidth, clip]{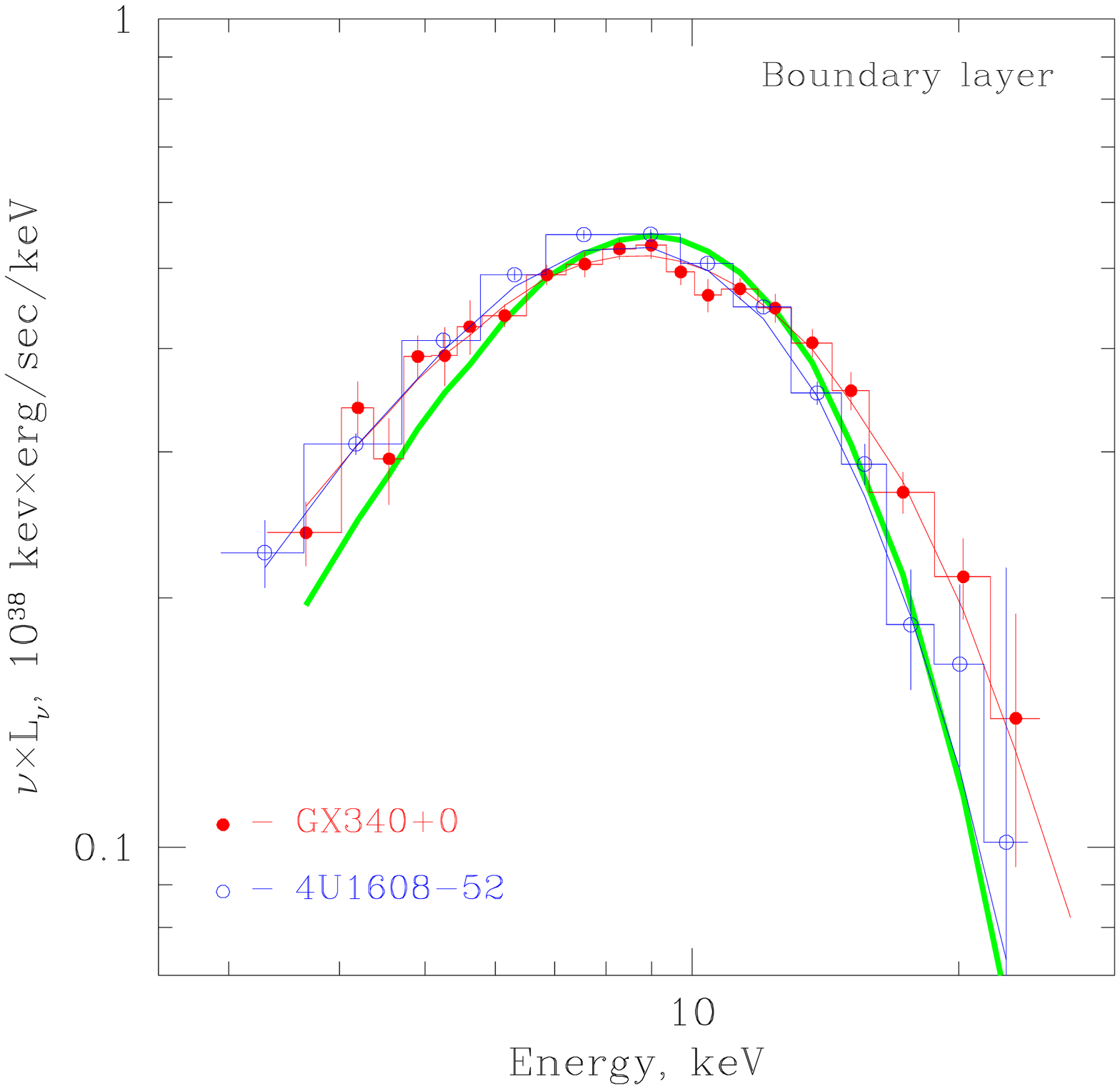}
\includegraphics[width=0.42\textwidth, clip]{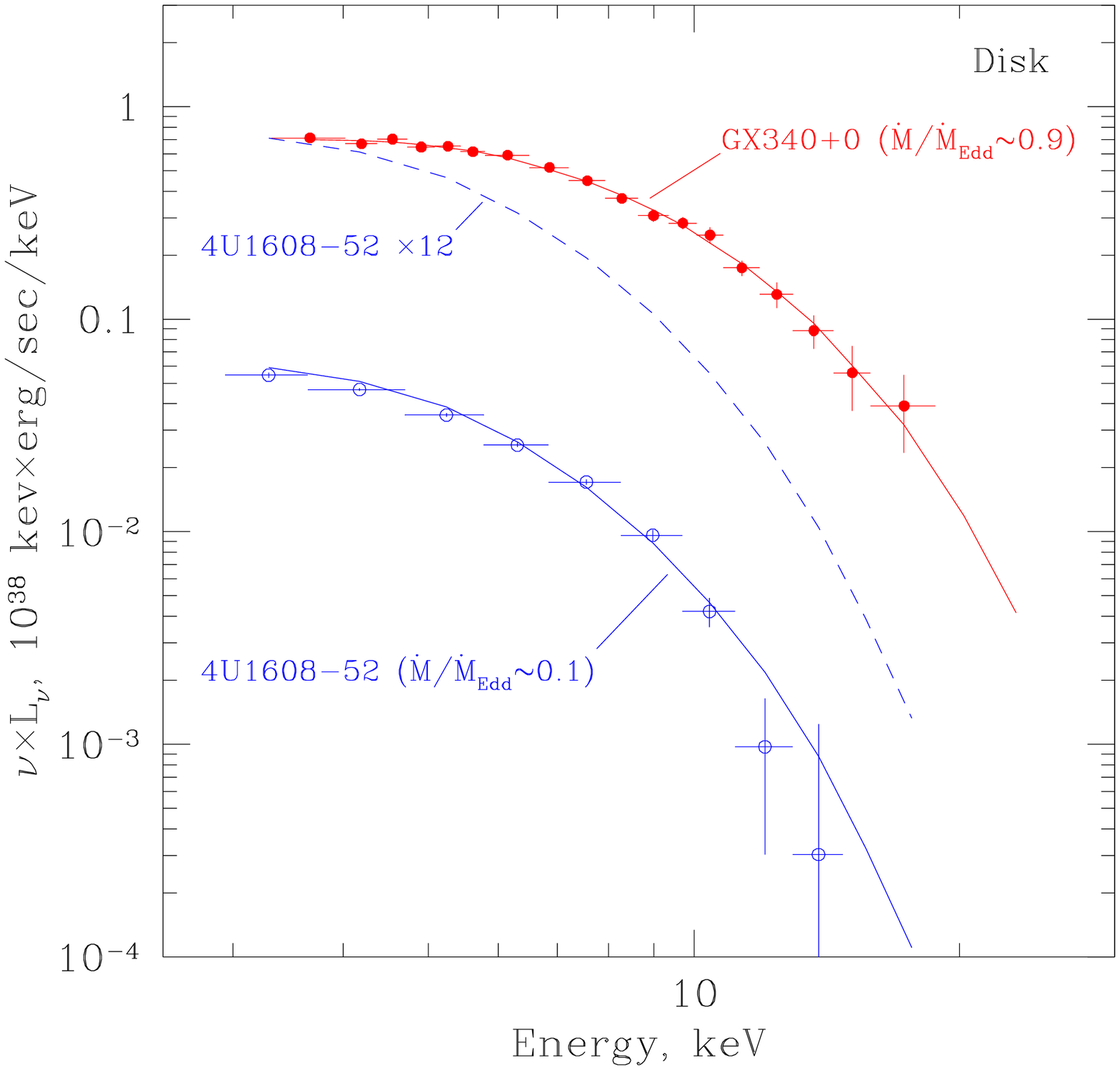}
}}
\caption{The absorption corrected average ({\em upper panel}),
frequency resolved ($\approx$ 
boundary layer, {\em middle}) and accretion disk ({\em lower})
spectra of GX340+0 and 4U1608--52. The solid, dashed and dotted lines
in the upper panel show the best fit total, boundary layer and
accretion disk spectra. The thin solid lines in the middle panel are best
fit Comptonized spectra, the thick grey line is Wien spectrum 
with $kt=2.1$ keV. The frequency resolved spectra
are normalized to the same energy flux. The
solid lines in the lower panel show best fit spectra of the accretion
disk, the dashed line is  4U1608--52 spectrum multiplied by a factor
of 12. See Table \ref{tb:fit} for values of the spectral parameters.
\label{fig:disk_bl}}
\end{figure}

The simplest models of the blackbody or Wien spectrum with $kT\sim
2-2.2$ keV, although they reproduce approximately the shape of the
frequency resolved spectra, show statistically significant  deviations 
from the data (Fig.~\ref{fig:disk_bl}) and give unacceptable value of
$\chi^2$. 
Significantly better the spectra can be described by the
Comptonized emission models \citep{st80,compps}. Although simple
Comptonization models are not expected to be directly applicable to
the emission spectra emerging from the boundary layer, we give in 
Table \ref{tb:fit}  the best fit parameters for comptt model in XSPEC
\citep{comptt} for easy comparison with  other results. 
A convenient parameterization of the frequency resolved spectra is
provided by a power law spectrum with exponential cut-off, which we
will use for the following analysis of the average spectra.

As the second step, we approximate the average spectra by a two
component model including the disk, boundary layer emission
and the reflected component appearing due to reflection of the latter
from the accretion disk: phabs$\times$(grad+pexrav+gaussian). 
The shape (but not the normalization) of the boundary layer emission
was fixed at the values determined from approximation of the
frequency resolved spectrum. The reflection was modeled using pexrav
model \citep{pexrav} and a broad gaussian line at 6.4--6.7 keV.
All parameters of the grad model were 
fixed except the mass accretion rate. The mass of the neutron star was
fixed at 1.4$M_{\sun}$. The disk inclination angle and the
color-to-effective temperature ratio were fixed at somewhat arbitrary,
although reasonable, values, which were chosen to
approximately  minimize the residuals. Note, that presence of the line
at $E\sim 6-7$ keV in the average spectrum is statistically
significatly required by the data,  
especially in the case of 4U1608--52. The parameters of the spectral
fits are summarized 
in the Table \ref{tb:fit}. The best fit models and different
components are depicted Fig.~\ref{fig:blspe} and \ref{fig:disk_bl}.
Although the $\chi^2$ values of the fits to the average spectra are in
the range 3.0--4.5 per d.o.f. and are formally unacceptable, the models
describe the observed total spectra (having very statistical
significance, signal/noise$\sim 10^2-10^3$ per energy channel)
reasonably well,  with relative accuracy of $\la 1-2\%$ in the 3--20
keV band.

\section{Discussion}
\label{sec:discussion}

The best fit values of the mass accretion rates obtained from
the spectral fits agree, within a factor of $\la 1.7$ with the values
predicted from the observed energy flux and the  accretion efficiency
expected for a neutron star with spin frequency of $\sim 500$ Hz
(Eq.~(\ref{eq:mdot})).
This fact is especially encouraging, as the two sources have accretion
rates different by a factor of $\sim 10$ (Table \ref{tb:fit}). For such
difference in the accretion rate the expected disk 
temperatures should differ by a factor for $\sim 1.7-1.8$, resulting in
the different shape and total flux of the disk spectra
(Fig.~\ref{fig:disk_bl}, lower panel). However, after the contribution
of the boundary  layer is accounted for, the relativistic disk emission 
model is capable of reproducing both spectral shape and normalization. 

Despite large difference in the mass accretion rate 
in the two sources, the energy spectra of the boundary layer emission are 
very similar to each other. 
This is in line with the finding that variations of the boundary 
layer luminosity in the broad range of time scales from  $\sim$sec
to $\sim$msec are not accompanied by  significant variations of the
spectral shape. Similar behavior was found by \citet{mitsuda84} and 
\citet{mitsuda86} on longer time scales of $\sim 10^3$ sec. 

Due to short light travel time of the accretion disk in the vicinity
of the neutron star, $\sim$msec, the reflected component, if originating
in the inner disk, could contribute to the variable emission
and cause  deviation of the frequency resolved spectra from the
true boundary layer spectrum. This can not be directly verified with
the present data --  upper limit on the equivalent width of the 6.4-6.7
keV line is in the case of both sources $\approx 110$ eV (90\%
confidence). However, for the observed shape of the spectrum of
the boundary layer, contribution of the reflected component with
$\Omega/2\pi\sim 0.2-0.3$, if any, would not exceed $10\%$ in the
$\sim 10-20$ keV energy range, i.e. is comparable or smaller than
other uncertainties involved.

\begin{figure}
\includegraphics[width=0.5\textwidth, clip]{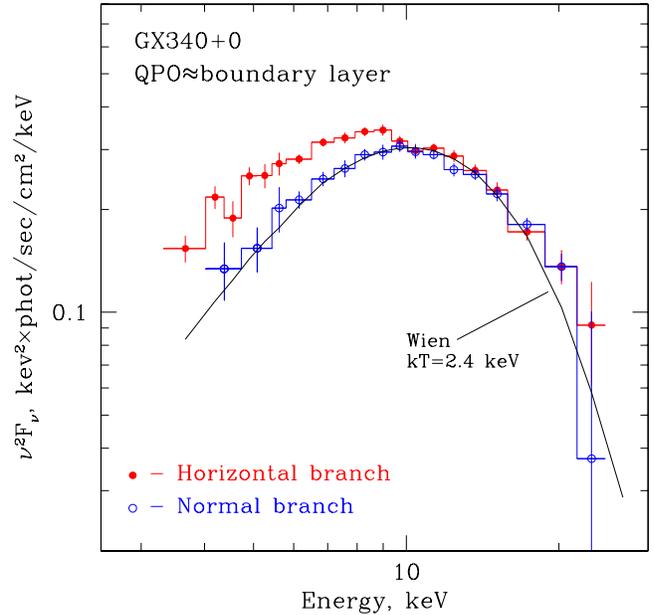}
\caption{The absorption corrected frequency resolved spectra of QPO
($\approx$ boundary layer emission) in GX340 in the horizontal branch,
Z=0--1 (lower $\dot{M}$) and upper half of the normal branch, Z=1--1.5
(higher $\dot{M}$). See Fig.~\ref{fig:cid} for specification of the
regions in the color-intensity diagram. The horizontal branch  
data is same as in Fig.~\ref{fig:freqres_gx340}, \ref{fig:blspe} and
\ref{fig:disk_bl}. The solid line shows  Wien spectrum with $kT=2.4$
keV.  
\label{fig:bl_alongz}}
\end{figure}

Further along the Z-track of GX340+0 in the color-intensity
diagram, on  normal and flaring branches,  the fractional rms of 
the X-ray variability decreases significantly, by a factor of $\sim
5-10$. However, the statistics  
is sufficient to place meaningful constrains on the behavior 
of the frequency resolved spectra at the first half of the normal
branch (Fig.~\ref{fig:cid}). 
The data indicates that the behavior of the  frequency resolved 
spectra does not change its character -- at sufficiently high frequency, 
$f\ga 1$ Hz, their shape does not depend upon 
the Fourier frequency and is significantly harder than the average spectrum 
and expected spectrum of the accretion disk. 
Therefore, in the same line of arguments as above, it is 
representative of the spectrum of the boundary layer spectrum. 
Fit to the frequency resolved spectrum by Comptonization model
requires infinitely large values of the  Comptonization parameter.
Correspondingly, the boundary layer spectrum in the normal branch  can
be well fit by Wien or blackbody spectrum (which are close to each
other in the 3-20 keV range)  with the best fit 
temperature of $kT\approx 2.4$ keV. 
As evident from Fig.~\ref{fig:disk_bl} and, especially, from
Fig.~\ref{fig:bl_alongz} the high energy part, E$\ga 8$ keV of the
spectrum of 4U1608--52 and horizontal branch of GX340+0
also follows Wien spectrum with temperature in
the range $kT\sim 2.1-2.3$.  
The composite fit of the total spectrum with the disk + boundary layer 
spectrum, the same as in Subsect.~\ref{sec:fit}, gives a best fit value 
of the mass accretion rate of $\dot{M}\approx 4.6 \cdot 10^{18}$ g/s,
i.e. higher than in the horizontal branch (cf. Table
\ref{tb:fit}). This is consistent with the commonly accepted  
interpretation that the mass accretion rate increases along the 
Z-track on the  color-intensity diagram.

The  frequency resolved spectra ($\approx$boundary layer spectra) on
the normal and horizontal branch of the color-intensity diagram are 
plotted in Fig.~\ref{fig:bl_alongz} along with Wien spectrum 
with $kT=2.4$ keV. Combined with the middle panel in
Fig.~\ref{fig:disk_bl} this plot shows trend in the dependence of the  
boundary layer spectrum upon the mass accretion rate in the range
$\dot{M}\sim (0.1-1.0) \dot{M}_{\rm Edd}$. We can tentatively
conclude that with increase of the mass accretion rate up to a value
close to critical Eddington rate the boundary layer spectrum in the
3--20 keV energy  range approaches a Wien spectrum. 
Interestingly, at lower values of $\dot{M}$
the character of the deviations of the boundary layer spectrum from the
Wien spectrum is similar to that expected in the  situation when
Compton scatterings are important factor of the spectral formation in
the media with inhomogeneous temperature distribution \citep{ross96}.  
In particular they are qualitatively  similar to the numerical results
of \citet{greb} on the  formation of the spectrum of the boundary layer.

The relatively weak dependence of the shape of the boundary layer spectrum   
upon the mass accretion rate (Fig.~\ref{fig:disk_bl} and
\ref{fig:bl_alongz}) and the relative constancy of the   
Wien temperature is somewhat surprising. It implies that in the
considered range of $\dot{M}\ga 0.1 \dot{M}_{\rm Edd}$ the
plasma temperature at Comptonization depth in the boundary layer
weakly depends upon the mass accretion rate, i.e. increase of the
$\dot{M}$ does not change significantly vertical temperature structure
in the boundary layer.

The fact that kHz QPO show the same behavior as other components
of the aperiodic variability indicates, that they have the same origin,
i.e. are caused by the variations of the luminosity of the boundary
layer. Although the kHz ``clock'' can be in the disk or due to it's
interaction with the neutron star, the actual modulation of the X-ray
flux  occurs on the neutron star surface. \citet{mendez0614} suggested
similar interpretation of kHz QPO in 4U0614+09. In particular they
showed, that  energy spectrum of the kHz 
QPO can be approximately described by a blackbody spectrum with
$kT\sim 1.5-1.6$ keV. Note, however, that they found different energy
depedence of continuum aperiodic variability at lower
frequencies. That can possibly be explained by the fact that the
source was in significantly lower luminosity state, $\dot{M}\sim 
10^{-2}\dot{M}_{\rm Edd}$ and it's energy spectrum had  
a distinct power law component which could dominate variability at
lower frequencies. 

\begin{figure}
\includegraphics[width=0.5\textwidth, clip]{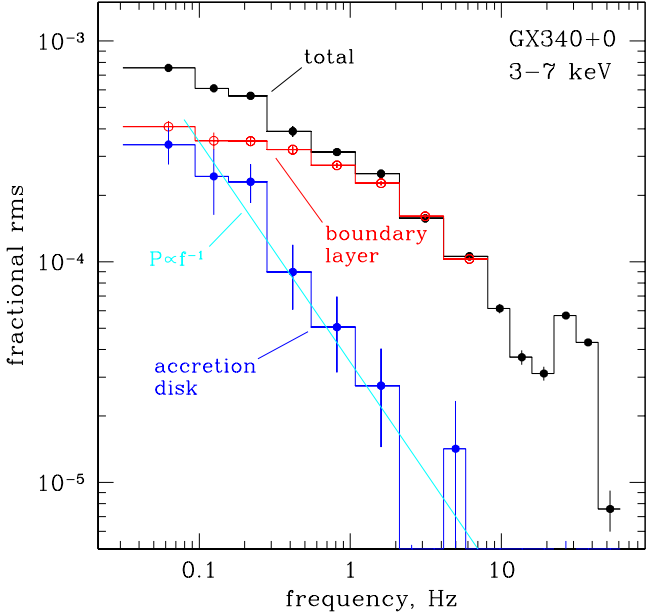}
\caption{The  power density spectra of GX340+0 in the 3--7 keV energy
range (horizontal branch, same data as in Fig.~\ref{fig:freqres_gx340},
\ref{fig:blspe} and \ref{fig:disk_bl}). The histogram with the error
bars show power spectra of total, disk and boundary layer emission,
obtained using the  procedure described in the
Sect.~\ref{sec:discussion}; the straight  
solid line shows a power law $P(f)=3.5e-5\times f^{-1}$. 
The power density is shown in units of fractional rms per Hz and all
three spectra are normalized to the total count rate in the 3--7 keV
band. Note, that the lower-most frequency bin is affected by the
windowing effects, suppressing the power. 
\label{fig:pds_disk_bl}}
\end{figure}

The disk emission is significantly less variable and does not
contribute significantly to the variability of the X-ray flux at 
$f\ga 0.5-1$ Hz. 
This conclusion is in agreement with results of 
\citet{chur01} for the high state of Cyg X-1, indicating that
stability of the X-ray emission might be a common property of the
optically thick accretion disk, independently on the nature of the compact
object. 
The origin of the variable component, however, is different in the
case of Cyg X-1 (and presumably in the soft state of other black hole
binaries).    
Indeed, the spectrum of the variable component in the soft state of
Cyg X-1 is identical to the 
time average spectrum of the hard spectral component and is
adequatly represented by unsaturated Comptonization in hot ($kT_{\rm
e}\sim 50-100$ keV) and optically thin ($\tau_{\rm T}\la 1$) coronal
flow with possible contribution of non-thermal
Comptonization. Variable component in luminous LMXBs considered in
this paper, if interpreted in the framework of the Comptonization
model,  requires saturated Comptonizaion (Comptonization parameter
$y\sim 1$) in the relatively low temperature ($kT_{\rm e}\sim 2-3$ keV)
plasma and is inconsistent both qualitatively and quantitatively  with
the corona models usually applied to black hole binaries.

In the bright LMXB systems, the contribution of the disk
variability becomes noticeable at  lower frequencies, below
$\sim 0.5$ Hz, where the frequency resolved spectrum changes it's
shape and becomes softer  
(Fig.~\ref{fig:freqres_gx340} and \ref{fig:freqres_gx340_ratios};
cf. results of \citealt{vdk86} for GX5--1).  
This can be used to estimate the contribution of the disk to the
observed variability of the X-ray flux. The result depends on the
character  of the disk variability. In order to make a crude estimate
we assume that the disk variations also obey a simple linear
relation described by Eq.~(\ref{eq:lc}):
\begin{eqnarray}
F(e,t)=F_{\rm disk}(E,t)+F_{\rm BL}(E,t)\approx\\
\approx S_{\rm disk}(E)\times f_{\rm disk}(t)+
S_{\rm BL}(E)\times f_{\rm BL}(t) \nonumber
\label{eq:lc_disk_bl}
\end{eqnarray}
If disk and boundary layer variations were uncorrelated, the power
density of the total signal would be:
\begin{eqnarray}
P(E,\omega)\propto S_{\rm disk}(E)^2 \times |\hat{f}_{\rm disk} (\omega)|^2+
S_{\rm BL}(E)^2 \times |\hat{f}_{\rm BL} (\omega)|^2 \nonumber
\label{eq:pds_disk_bl}
\end{eqnarray}
where within the accuracy of this consideration one can assume that
$S_{\rm disk}(E)$ and $S_{\rm BL}(E)$ are the disk and boundary layer
spectra determined in Sect.~\ref{sec:fit}. The functions
$|\hat{f}_{\rm disk} (\omega)|^2$ and  $|\hat{f}_{\rm BL} (\omega)|^2$
after appropriate renormalization represent power density spectra of
the disk and boundary layer and can be determined from linear fit to
the square of the frequency resolved spectra in each frequency
interval.  The power density spectra thus computed are shown in
Fig.~\ref{fig:pds_disk_bl} along with the total 
power spectrum  of GX340+0 in the soft 3--7 keV energy band.
The power spectrum of the accretion disk flux variations is different
from that of the boundary layer, does not extend significantly to
the high frequency domain, and is consistent with a power law with
slope of $-1$:
$$
P_{\rm disk}(f)\propto f^{-1}
$$ 
The excess power seen at low frequencies, $F\la 0.5-1$ Hz in the
soft energy band (cf. Fig.~\ref{fig:pds_gx340}) can be explained as the
contribution of the disk variations. At higher frequencies the
variability is dominated by the boundary layer emission, giving
primary contribution to quasi-periodic oscillations  and the so called
band limited noise compoinent \citep[e.g.][]{vdk86}.

\section{Summary}
\label{sec:summary}

The initial observational results are listed in
Subsect.~\ref{sec:obs_sum}. Below we summarize the constrains on the
character of the spectral variability and implications for the
boundary layer and accretion disk models.

\subsection{Constrains on the pattern and origin of the spectral 
variability in luminous LMXBs}

\begin{enumerate}

\item
Using RXTE/PCA observations of two luminous low mass X-ray binaries
GX340+0 (on the horizontal/normal branch of the color-intensity
diagram) and 4U1608-52 we show that the shape of the  Fourier
frequency resolved spectra 
on $\sim$ second -- millisecond time scales does not depend on Fourier 
frequency (Fig.~\ref{fig:freqres_gx340}, \ref{fig:freqres_gx340_ratios} 
and \ref{fig:freqres_1608}). \
The range of investigated  timescales includes the band limited
continuum noise,  the kHz QPO and lower 
frequency QPOs observed at few tens Hz 
(Fig.~\ref{fig:pds_gx340}, \ref{fig:pds_1608}). 
Combined with the negligibly small
phase lags, $\Delta\phi\la 10^{-2}$ (Fig.~\ref{fig:lags_gx340}),  
this restricts significantly the possible pattern of spectral 
variability of X-ray 
flux and requires linear relation between flux variations at different 
energies (Eq.~(\ref{eq:lc})).

Considering significant difference in the expected spectra of 
the accretion disk and boundary layer the observed variations should
be associated with either one of these two major components 
of the accretion flow.
The X-ray variability is caused either by variations of 
it's luminosity under constant spectral shape, or 
by small variations of a spectral parameter 
(e.g temperature or optical depth) -- Eq.~(\ref{eq:taylor}).

\item 
We compared the 
Fourier frequency resolved spectra with the  expected 
spectra of the accretion disk and of the boundary layer. The predicted
spectra were based on the observed energy flux/spectrum and 
very generic system parameters such as the source distance and
neutron star spin frequency. 
The frequency resolved spectra are well consistent with the  
range of the boundary layer spectra expected for plausible 
range of the system parameters (Fig.~\ref{fig:blspe}, Table
\ref{tb:disk_range}). 
On the other hand, they are significantly harder than the expected
spectrum of the accretion disk. It is unlikely that the observed
variations are associated with variations of the disk luminosity or 
spectral shape unless the disk temperature is $\sim 3-4 $ kev, i.e. current 
accretion disk models are inapplicable to the neutron star 
binaries.

\item
The above suggests that the major part of aperiodic and quasiperiodic 
variability observed in luminous LMXBs above $\sim 0.5$ Hz
is caused by variations of the luminosity of the boundary layer. 
Its spectral shape remains nearly constant in the course of the 
luminosity variations. This interpretations receives additional 
support from the constancy of the fractional rms with energy 
at $E\ga 10$ keV, where expected accretion disk emission vanishes, 
found in case of 4U1608--52 (Fig.~\ref{fig:rms_1608}).

\end{enumerate}

\subsection{Implications for the models of the boundary layer and 
disk emission}

The frequency resolved spectrum is representative
of the energy spectrum of the 
boundary layer emission. This can be used for a more precise decomposition 
of the spectra of luminous LMXBs into accretion disk and boundary 
layer components and for quantitative comparison with  predictions 
of the theoretical models. In the following we shall assume that
boundary layer spectrum is identical to the  frequency resolved
spectrum, bearing in mind that this is true to certain accuracy.

\begin{enumerate}

\item
In the considered range of the mass accretion rate $\dot{M}\sim
(0.1-1)\dot{M}_{\rm Edd}$, the boundary layer spectrum in the 3--20 keV
energy range  depends weakly on $\dot{M}$. 
Its shape is remarkably similar in GX340+0 and 4U1608--52
(Fig.~\ref{fig:disk_bl}), despite the fact that the two sources have a  
factor of $\sim10$ difference in the mass accretion rate (Table
\ref{tb:fit}). 

In the limit of high $\dot{M}$, of the order of 
$\sim\dot{M}_{\rm Edd}$ (normal branch of GX340+0), the boundary layer 
spectrum in the 3--20 keV  
energy range  can be adequately  represented by the Wien spectrum with 
temperature  $kT\approx 2.4$ keV (Fig.~\ref{fig:bl_alongz}). 
At lower values of $\dot{M}$ (4U1608--52 and horizontal branch of
GX340+0) the spectra are better described by Comptonization model with
electron temperature of $\sim 2-4$ keV and 
Comptonization parameter $y\sim 1$ (Table \ref{tb:fit}). Their high
energy part, $E\ga 10$ keV, is well represented by Wien spectrum with
temperature of $\approx 2.1-2.3$ keV.

\item
The average spectra can be adequately described by the sum of the
renormalized frequency resolved spectrum and the accretion disk emission
(Fig.~\ref{fig:disk_bl}). The spectrum of the latter is well described
by the general relativistic accretion disk model. 
The other parameters, such as source distance and disk inclination   
angle being fixed at fudicial but plausible values, the best fit value 
of the mass accretion rate coincides, within a factor of $\la 1.7$ with 
that inferred from the observed X-ray flux and accretion efficiency 
appropriate for a 1.4$M_{\sun}$ neutron star with spin frequency 
of $\sim 500$ Hz (Table \ref{tb:fit}).  
This agreement is especially remarkable, given the luminosity and mass
accretion rate in the two sources differ by the factor of $\sim 10$.

\item 
The accretion disk emission is significantly less variable than
the boundary layer emission at Fourier frequencies $f\ga 0.5-1$ Hz.
The power density spectrum of the disk appears to follow a power law  
$P_{\rm disk}(f)\propto f^{-1}$ and contributes to the overall
variability in the soft energy band and in the low frequency domain
only (Fig.~\ref{fig:pds_disk_bl}). 

\item
The  kHz QPOs apear to have the same origin as aperiodic and
quasiperiodic variability at lower  
frequencies. The msec flux modulations originate on the surface of the
neutron star although the kHz ``clock''  might reside in the disk or
be determined by the  disk -- neutron star interaction.  

\item 
Finally we point out that in the case of GX340+0 and presumably other
Z--sources, the above results apply to the normal and horizontal
branches of the color-intensity diagram. The source behaviour on the
flaring branch, believed to correspond to super-Eddington accretion is
more complex and is beyond the scope of this paper.

\end{enumerate}

\begin{acknowledgements}
The authors would like to thank Rashid Sunyaev, Nail Inogamov, Eugene
Churazov and Mariano Mendez for useful discussions. We are thankful
to the referee, Michiel van der Klis, for stimulating comments
which helped to improve the paper. S.M. acknowledges partial support 
by RFBR grant 03-02-06772.
This research has made use of data obtained through the High Energy
Astrophysics Science Archive Research Center Online Service, provided 
by the NASA/Goddard Space Flight Center. 
\end{acknowledgements}

\end{document}